\documentclass[12pt]{article}
\usepackage[english]{babel}
\usepackage[utf8]{inputenc}
\usepackage[left=2cm,right=2cm,top=2cm,bottom=2cm,bindingoffset=0cm]{geometry}
\usepackage{hyperref, xcolor, graphicx, authblk}
\usepackage[sort&compress,numbers]{natbib}
\usepackage{amsfonts, amssymb, amsmath, mathtools}

\newcommand\beq{\begin{equation}}
\newcommand\eeq{\end{equation}}
\newcommand\bem{\begin{pmatrix}}
\newcommand\eem{\end{pmatrix}}
\newcommand{\pd}{\partial}
\newcommand{\mO}{\mathcal{O}}

\newcommand{\mK}{\mathcal{K}}
\newcommand{\mT}{\hat{\mathrm{T}}}
\newcommand{\const}{\text{const}}
\newcommand{\tr}{\text{tr}}

\numberwithin{equation}{section}

\hypersetup{
    colorlinks=true,
    linkcolor=black,
    citecolor=blue,   
    urlcolor=blue,
}

\title{\bf Refined quantum Lyapunov exponents from replica out-of-time-order correlators}

\author[1,2]{Dmitrii~A.~Trunin\thanks{\href{mailto:dmitriy.trunin@phystech.edu}{dmitriy.trunin@phystech.edu}}}
\affil[1]{Moscow Institute of Physics and Technology, 141701, Institutskiy pereulok 9, Dolgoprudny, Russia}
\affil[2]{Lebedev Physical Institute, 119991, Leninskiy prospect 53, Moscow, Russia}

\date{\today}

\begin{document}


\maketitle

\begin{abstract}
We suggest a new indicator of quantum chaos based on the logarithmic out-of-time-order correlator. On the one hand, this indicator correctly reproduces the average classical Lyapunov exponent in the semiclassical limit and directly links the definitions of quantum chaos and classical K-system. On the other hand, it can be analytically calculated using the replica trick and the Schwinger-Keldysh diagram technique on a $2n$-fold Keldysh contour. To illustrate this approach, we consider several one-dimensional systems, including the quantum cat map, and three paradigmatic large-$N$ models, including the Sachdev-Ye-Kitaev model. Furthermore, we find that correlations between replicas can reduce the magnitude of the Lyapunov exponent compared to estimates based on conventional out-of-time-order correlators.
\end{abstract}

\tableofcontents

\newpage

\section{Introduction}
\label{sec:intro}

Chaotic behavior is ubiquitous in classical dynamical systems, justifies the use of macroscopic equilibrium variables in classical thermodynamics and hydrodynamics, and has been extensively studied since its
discovery in the early 1960s~\cite{Lorenz-1, Lorenz-2, Tabor, Ott}. As a rule, classical chaos is associated with hyperbolicity --- the exponential sensitivity to infinitesimal changes in initial conditions measured by a positive largest Lyapunov exponent (LE) and widely known as the butterfly effect:
\beq \label{eq:cl-LE-1}
\kappa_{cl}(\mathbf{z}_0) = \lim_{t \to \infty} \limsup_{\| \delta \mathbf{z} \| \to 0} \frac{1}{t} \log \frac{\| \delta \mathbf{z}(t; \mathbf{z}_0) \|}{\| \delta \mathbf{z} \|} > 0. \eeq
Here, $\mathbf{z} = (\mathbf{q}, \mathbf{p})$ are canonical coordinates, $\| \cdots \|$ is a norm on the phase space, and $\delta \mathbf{z}(t; \mathbf{z}_0)$ is the distance between two phase trajectories emanating at the moment $t = 0$ from two points close to~$\mathbf{z}_0$ but separated by a small initial distance $\delta \mathbf{z}$. It is straightforward to show that definition~\eqref{eq:cl-LE-1} does not depend on the choice of the norm (at least if the phase space can be equipped with a metric). Hence, we can choose a convenient Euclidean norm, explicitly take the limit $\| \delta \mathbf{z} \| \to 0$ and rewrite definition~\eqref{eq:cl-LE-1}:
\beq \label{eq:cl-LE-2}
\kappa_{cl}(\mathbf{z}_0) = \lim_{t \to \infty} \frac{1}{2t} \log \sum_{i,j} \Phi_{ji}^*(t; \mathbf{z}_0) \Phi_{ij}(t; \mathbf{z}_0), \eeq
where matrix $\Phi_{ij}(t; \mathbf{z}_0) = \pd z_i(t; \mathbf{z}_0) / \pd z_{0j}$ measures the stretching and rotation of an infinitesimal phase volume along the trajectories. The definition of the largest LE~\eqref{eq:cl-LE-2} is also generalized to cover the entire Lyapunov spectrum, which is encoded in eigenvalues of (diagonal) matrix $\Lambda_{ij}(\mathbf{z}_0)$:
\beq \label{eq:cl-LE-spectrum}
\mathbf{\Lambda}(\mathbf{z}_0) = \lim_{t \to \infty} \frac{1}{2t} \log \left[ \mathbf{\Phi}^\dag(t; \mathbf{z}_0) \mathbf{\Phi}(t; \mathbf{z}_0) \right]. \eeq

It is important to emphasize that definitions~\eqref{eq:cl-LE-1}--\eqref{eq:cl-LE-spectrum} depend on the starting point $\mathbf{z}_0$ and cannot distinguish between \textit{globally} and \textit{locally} hyperbolic systems. In fact, they yield positive LEs near an isolated saddle point even if the system is integrable, such as a double-well oscillator or simple pendulum. However, in such systems, LEs are positive only for a very small fraction of initial conditions, and their trajectories can be expressed in explicit functional form. Therefore, it is unnatural to classify them as chaotic systems. This shortcoming is corrected by averaging definition~\eqref{eq:cl-LE-1} or~\eqref{eq:cl-LE-2}:
\beq \label{eq:cl-LE-3}
\bar{\kappa}_{cl} = \int d \mathbf{z}_0 \, \mu(\mathbf{z}_0) \kappa_{cl}(\mathbf{z}_0), \eeq
where $\mu(\mathbf{z}_0)$ is a measure on the phase space, e.g., the probability density function of a thermal ensemble, $\mu(\mathbf{z}_0) \sim \exp\left[-\beta H(\mathbf{z}_0) \right]$. In a truly chaotic (i.e., globally hyperbolic) system, most trajectories exponentially diverge, so $\bar{\kappa}_{cl} > 0$. On the contrary, in an integrable system, most trajectories diverge in a power-law manner, and isolated saddle points do not contribute to the averaging, so $\bar{\kappa}_{cl} = 0$. Therefore, we call a classical system chaotic \textit{if and only if} $\bar{\kappa}_{cl} > 0$. In fact, this definition is equivalent to the definition of a K-system, which has a positive Kolmogorov-Sinai entropy $\kappa_{KS}$ \cite{Tabor, Ott, Pesin, Benettin}:
\beq \label{eq:KS-entropy}
\kappa_{KS} = \int d \mathbf{z}_0 \, \mu(\mathbf{z}_0) \sum_i \Lambda_{ii}(\mathbf{z}_0) \, \theta\left[ \Lambda_{ii}(\mathbf{z}_0) \right]. \eeq

Unlike the classical chaos, its quantum counterpart cannot be defined directly. There are two main obstacles that prevent an easy generalization of the definitions $\bar{\kappa}_{cl} > 0$ or $\kappa_{KS} > 0$ to the quantum case: the Heisenberg's uncertainty principle and the inherent unitarity of quantum mechanics. The first obstacle makes it impossible to define quantum chaos and quantum LEs by the exponential divergence of initially close \textit{trajectories}: in quantum mechanics, the phase space is not smooth, and, generally speaking, there are no trajectories. The second obstacle nips in the bud the idea of tracking the divergence of initially close solutions of the Schr\"{o}dinger equation, i.e., \textit{quantum states}: due to unitarity, the overlap of any two states remains constant even after a long evolution. Yet, the correspondence principle supposes the existence of quantum chaos, at least in the semiclassical limit. In other words, we expect a classically chaotic system to become quantum chaotic upon quantization (and the same for integrable systems). Therefore, we need to establish the key features of quantum chaos and invent an alternative indicator that distinguishes between chaotic and integrable quantum systems. 

Over the past forty years, there have been numerous attempts to capture the essence of quantum chaos and establish an indicator that is well-defined in quantum systems and distinguishes between chaotic and integrable classical systems upon their quantization. To date, there are about ten such indicators. The oldest and most famous one is the statistics of energy level spacings, which was suggested in the early eighties~\cite{Haake, Stockmann, Casati, Berry, Bohigas}. More recent examples of a quantum chaos indicator include the dynamical entropy~\cite{Connes, Alicki}, Loschmidt echo~\cite{Goussev, Jalabert:2001}, decoherence~\cite{Zurek}, entanglement~\cite{Nie, Alba}, spectral form factor~\cite{Cotler, deMelloKoch, Gharibyan}, Hilbert-space geometry~\cite{Kolodrubetz, Pandey, Orlov}, out-of-time-order correlation functions~\cite{Larkin, MSS, Kitaev-talks, Swingle-popular} and their generalizations~\cite{Kukuljan, Lewis-Swan}, circuit~\cite{Ali, Bhattacharyya:2021} and Krylov~\cite{Parker:2018, Avdoshkin, Gorsky, Rabinovici, Smolkin, Caputa, Bhattacharjee:2022} complexities. Furthermore, some of these indicators are related, which suggests the existence of a broad ``web of quantum chaos indicators''~\cite{Bhattacharyya:2019, Kudler-Flam}.

Out-of-time-order correlator (OTOC) is probably the most remarkable indicator of quantum chaos. First of all, it naturally generalizes the concept of the LE to quantum systems, which makes the correspondence between the quantum and classical chaos explicit:
\beq \label{eq:OTOC-def}
C(t) = \frac{1}{N \hbar^2} \sum_{i,j} \left\langle \left[ \hat{z}_i(t), \hat{z}_j(0) \right]^\dag \left[ \hat{z}_i(t), \hat{z}_j(0) \right] \right\rangle \sim \sum_{i,j} \left\langle \Phi^*_{ji}(t; \mathbf{z}_0) \Phi_{ij}(t; \mathbf{z}_0) \right\rangle \sim \left\langle e^{2 \kappa_{cl} t} \right\rangle \sim e^{2 \kappa_q t}. \eeq
Here, $N$ is the phase space dimension, $\langle \, \cdots \rangle$ denotes the averaging over a suitable quantum or classical ensemble (usually a thermal one), and $\kappa_q$ is referred to as the (conventional) quantum LE. In derivation of Eq.~\eqref{eq:OTOC-def}, we use the relation between the commutator and the Poisson bracket in the semiclassical limit: $\Phi_{ij}(t; \mathbf{z}_0) = \left\{ z_i(t), z_j(0) \right\} \sim \frac{1}{i \hbar} \left[ \hat{z}_i(t), \hat{z}_j(0) \right]$ as $\hbar \to 0$. Thus, we can define a quantum chaotic system as a system with $\kappa_q > 0$. Roughly speaking, the OTOC tracks the traces of the semiclassical trajectories in a quantized system, and $\kappa_q$ measures the exponential divergence of such would-be trajectories.

Note, however, that the semiclassical approximation used for the derivation of Eq.~\eqref{eq:OTOC-def} is valid only well before the Ehrenfest time $t_E$, where all initially narrow wave packets are substantially smeared~\cite{Shepelyansky:Scholarpedia, Zaslavsky, Chirikov-88, Aleiner-96}. This time scales with $\hbar$ polynomially in an integrable system, $t_E \sim 1/\hbar^{N/2}$, and logarithmically in a chaotic system, $t_E \sim \log(1/\hbar)$. Therefore, the exponential growth of the OTOC cannot continue forever, and the quantum LE is extracted only from this short interval:
\beq \label{eq:naive-qLE-def}
\log\!\left[ C(t) \right] = 2 \kappa_q t + o(t) \quad \text{as} \quad 1 \ll t \ll t_E, \eeq
where the function $o(t)$ grows slower than linearly, i.e., $o(t)/t \to 0$ as $t \to \infty$. Note that the behavior of the OTOC at larger times also contains some information about the properties of a quantum chaotic system, even though it cannot be interpreted using the semiclassical picture~\cite{Markovic}.

Besides, OTOCs have several other advantages. First, they are easily generalized to arbitrary quantum systems, including many-body systems, by promoting operators $\hat{z}_i$ to any other suitable local operators. Second, OTOCs can be used as a crude measure of operator growth and scrambling, i.e., propagation of information about initial perturbations throughout a quantum system~\cite{Swingle:2016, Roberts:2016, Nahum, Mi, Xu:2018, Xu-tutorial, Shenker-1, Shenker-2, Roberts-1, Shenker-3}. Third, quantum LE defined by~\eqref{eq:naive-qLE-def} is believed to satisfy the bound $\kappa_q \le 2 \pi T/\hbar$ and saturate it for black holes; hence, the saturation of this bound by a quantum system hints that it might be holographically dual to a black hole and serve as a qualitative model of its microstates~\cite{MSS}. Finally and most importantly, OTOCs are relatively easy to calculate --- unlike most other indicators of quantum chaos. All these advantages make the OTOC a very popular tool. To date, the OTOCs were calculated in a vast number of models, including the Ba\~{n}ados-Teitelboim-Zanelli black hole~\cite{Shenker-1, Shenker-2, Roberts-1, Shenker-3} and its dual conformal field theory~\cite{Roberts-2, Fitzpatrick, Turiaci}, Sachdev-Ye-Kitaev model~\cite{Polchinski, Maldacena-SYK, Kitaev, Sarosi, Rosenhaus, Trunin-SYK} and Jackiw-Teitelboim gravity~\cite{Maldacena-JT, Jensen, Engelsoy}, various quantum mechanical models~\cite{Hashimoto:2017, Akutagawa, Romatschke, Michel, Buividovich:2018, Buividovich:2022, Kolganov}, quantum many-body systems~\cite{Hosur, Huang, Shen, Mezei, Dora, Bohrdt, Patel, Klug, Lin, Tikhanovskaya, Kent:2023}, and quantum field theories~\cite{Stanford:phi-4, Grozdanov, Romero-Bermudez, Liao, Sahu, Swingle:2017, Steinberg, Anninos, Aalsma}. Moreover, OTOCs were experimentally measured in several prominent systems~\cite{Garttner, Li, Green}, including the model of a traversable wormhole~\cite{Jafferis:Nature}.

Unfortunately, OTOCs were recently shown to mistake quantized integrable systems for chaotic ones, thus violating the expected correspondence between classical and quantum chaos~\cite{Rozenbaum-1, Rozenbaum-2, Hashimoto:2020, Xu, Pilatowsky-Cameo:2019, Wang:2018, Pappalardi:2018, Hummel:2018, Steinhuber:2023, Dowling:2023}. The origin of this discrepancy is as follows. In the definition of the classical LE, Eq.~\eqref{eq:cl-LE-3}, we first take the logarithm of sensitivity, Eq.~\eqref{eq:cl-LE-2}, and then average it over the phase space to suppress the contributions of isolated saddle points. However, in the definition of the quantum LE, Eq.~\eqref{eq:naive-qLE-def}, the order of taking logarithm and averaging is different, so the contribution of isolated saddle points is not suppressed~\cite{Xu}. To illustrate this problem, let us consider a simple example of an integrable system with an isolated saddle point, e.g., a double-well oscillator:
\beq \label{eq:double-well-H}
H = \frac{1}{2} p^2 + \frac{1}{4} (1 - q^2)^2, \eeq
where, for simplicity, we take all parameters to be dimensionless. The Hamiltonian equations of motion imply the saddle point $(q,p) = (0,0)$:
\beq \label{eq:double-well-saddle}
\begin{cases} \dot{q} = p, \\ \dot{p} = q - q^3 \approx q, \end{cases} \Longrightarrow \quad \begin{cases} q(t) + p(t) \approx (q_0 + p_0) e^t, \\ q(t) - p(t) \approx (q_0 - p_0) e^{-t}. \end{cases} \eeq
If the initial point belongs to a narrow strip $\mathcal{S}_{t_*} = \{ |q_0 + p_0| < \delta e^{-t_*}, |q_0 - p_0| < \delta \}$, where $\delta \ll 1$, the exponential growth~\eqref{eq:double-well-saddle} persists until $t \approx t_*$. Hence, we can estimate the matrix $\Phi_{ij}(t; \mathbf{z}_0)$:
\beq \label{eq:double-well-sensitivity}
\mathbf{\Phi}(t; \mathbf{z}_0) \approx \bem \cosh t & \sinh t \\ -\sinh t & \cosh t \eem \quad \text{for} \quad \mathbf{z}_0 \in \mathcal{S}_t. \eeq
Now, we substitute this expression into the semiclassical approximation of the OTOC, Eq.~\eqref{eq:OTOC-def}, assume the averaging over a thermal ensemble, and keep in mind that $\Phi_{ij}(t; \mathbf{z}_0)$ does not grow exponentially outside the strip $\mathcal{S}_t$ due to integrability of the model~\eqref{eq:double-well-H}. Thus we estimate the OTOC in the quantized version of model~\eqref{eq:double-well-H}:
\beq \label{eq:double-well-OTOC}
C(t) \sim \int_{\mathcal{S}_t} d\mathbf{z}_0 \sum_{i,j} \Phi_{ji}^*(t; \mathbf{z}_0) \Phi_{ij}(t; \mathbf{z}_0) e^{-\beta H(\mathbf{z}_0)} \sim \mathrm{vol}(\mathcal{S}_t) \times \cosh^2(t) \sim \delta^2 e^{-t} \times e^{2 t} \sim \delta^2 e^t. \eeq
Hence, the quantum LE $\kappa_q = 1/2$, whereas the average classical LE $\bar{\kappa}_{cl} = 0$. In fact, it is straightforward to show that $\kappa_q \to \max\left[\kappa_{cl}(\mathbf{z}_0)\right]$ as $\hbar \to 0$, where the maximum is taken over the entire phase space~\cite{Kolganov, companion}. Evidently, the definition of quantum chaos based on $\kappa_q > 0$ is not equivalent to the definition of classical chaos $\bar{\kappa}_{cl} > 0$ in the limit $\hbar \to 0$. This discrepancy is similarly proved for an arbitrary quantized integrable system with isolated saddle points~\cite{Xu}.

To fix this flaw in the definition of quantum chaos but retain all advantages of the OTOC, we introduce the logarithmic OTOC (LOTOC) $L(t)$:
\beq \label{eq:LOTOC-def}
L(t) = \left\langle \log\!\bigg( \frac{1}{N \hbar^2} \sum_{i,j} \left[ \hat{z}_i(t), \hat{z}_j(0) \right]^\dag \left[ \hat{z}_i(t), \hat{z}_j(0) \right] \bigg) \right\rangle, \eeq
and the refined quantum LE $\bar{\kappa}_q$:
\beq \label{eq:true-qLE-def}
L(t) = 2 \bar \kappa_q t + o(t) \quad \text{as} \quad 1 \ll t \ll t_E, \eeq
where the function $o(t)$ again grows slower than linearly. The correct order of averaging and taking logarithm\footnote{Note that the different order of averaging and taking logarithm in the definitions of the conventional quantum LE~\eqref{eq:naive-qLE-def} and the refined quantum LE~\eqref{eq:true-qLE-def} resembles the difference between the Landauer and Lyapunov exponents, which appear in the context of Anderson localization~\cite{Kramer, Anderson, Sedrakyan}.} in~\eqref{eq:LOTOC-def}, supported by case studies of~\cite{companion}, implies that the refined quantum LE~\eqref{eq:true-qLE-def} reproduces the average classical LE~\eqref{eq:cl-LE-3} in the semiclassical limit $\hbar \to 0$. This ensures the correspondence between the classical chaos defined by $\bar{\kappa}_{cl} > 0$ and the quantum chaos defined by $\bar{\kappa}_q > 0$.

In this paper, we study the definitions of LOTOC~\eqref{eq:LOTOC-def} and refined quantum LE~\eqref{eq:true-qLE-def} more carefully. Namely, we discuss how to calculate these quantities analytically using the replica trick:
\beq \label{eq:replica-trick}
L(t) = \lim_{n \to 0} \frac{\pd C_n(t)}{\pd n} \qquad \text{and} \qquad \bar{\kappa}_q = \lim_{n \to 0} \frac{\pd \kappa_n}{\pd n}, \eeq
where we introduce an auxiliary replica OTOC (ROTOC):
\beq \label{eq:ROTOC-def}
C_n(t) = \left\langle \bigg( \frac{1}{N \hbar^2} \sum_{i,j} \left[ \hat{z}_i(t), \hat{z}_j(0) \right]^\dag \left[ \hat{z}_i(t), \hat{z}_j(0) \right] \bigg)^n \right\rangle, \eeq
and the replica LE\footnote{In the semiclassical limit, replica exponents reproduce the generalized LEs defined from the moments of the sensitivity distribution~\cite{Crisanti}.} $\kappa_n$:
\beq \label{eq:rLE-def}
\log\!\left[ C_n(t) \right] = 2 \kappa_n t + o(t) \quad \text{as} \quad 1 \ll t \ll t_E. \eeq
In other words, we show that the LOTOC has the same crucial advantage as the OTOC: it is easy to calculate in all systems where OTOCs proved to be useful, including models~\cite{Shenker-1, Shenker-2, Roberts-1, Shenker-3, Roberts-2, Fitzpatrick, Turiaci, Polchinski, Maldacena-SYK, Kitaev, Sarosi, Rosenhaus, Trunin-SYK, Maldacena-JT, Jensen, Engelsoy, Hashimoto:2017, Akutagawa, Romatschke, Michel, Buividovich:2018, Buividovich:2022, Kolganov, Hosur, Huang, Shen, Mezei, Dora, Bohrdt, Patel, Klug, Lin, Tikhanovskaya, Kent:2023, Stanford:phi-4, Grozdanov, Romero-Bermudez, Liao, Sahu, Swingle:2017, Steinberg, Anninos, Aalsma, Garttner, Li, Green, Jafferis:Nature}.

This paper is organized as follows. In Sec.~\ref{sec:simple}, we calculate the ROTOCs, LOTOCs, and the refined quantum LEs in the simplest one-dimensional dynamical models: the ordinary harmonic oscillator and the Lipkin-Meshkov-Glick model. In the first model, we perform all calculations explicitly; in the last model, we estimate the ROTOCs by employing the semiclassical approximation and then compare this estimate with numerical results. In Sec.~\ref{sec:map}, we test our definition of quantum chaos on a quantum chaotic map (a quantized Arnold cat map). In Sec.~\ref{sec:diagrams}, we discuss how to calculate ROTOCs using the Schwinger-Keldysh technique on a multi-fold Keldysh contour. Namely, we introduce this technique in Sec.~\ref{sec:basics} and apply it to three prominent large-$N$ models: the nonlinear vector mechanics with explicitly broken $O(N)$ symmetry (Sec.~\ref{sec:ON}), the Sachdev-Ye-Kitaev model (Sec.~\ref{sec:SYK}), and a weakly coupled matrix field theory (Sec.~\ref{sec:matrix}). Finally, we discuss the results and outline possible directions for future work in Sec.~\ref{sec:discussion}.

\section{Simple one-dimensional examples}
\label{sec:simple}

\subsection{Harmonic oscillator}
\label{sec:oscillator}

We start with the simplest textbook example, the quantum Harmonic oscillator:
\beq \label{eq:harmonic-H}
\hat{H} = \frac{1}{2} \hat{p}^2 + \frac{1}{2} \hat{q}^2. \eeq
For convenience, we measure the momentum and energy in units of length, $p \to m \omega p$, $H \to m \omega^2 H$, where $m$ and $\omega$ are the mass and the frequency of the original oscillator. Then we rewrite the Hamiltonian~\eqref{eq:harmonic-H} using the ladder operators $[\hat{a}, \hat{a}^\dag] = 1$ (note that the Planck's constant is also measured in units of length, $\hbar \to m \omega \hbar$):
\beq \label{eq:harmonic-H-ladder}
\hat{H} = \hbar \left( \hat{a}^\dag \hat{a} + \frac{1}{2} \right). \eeq
Here, we use the decomposition of the position and momentum operators at the moment $t = 0$:
\beq \label{eq:harmonic-qp-0}
\hat{q}(0) = \sqrt{\frac{\hbar}{2}} \left( \hat{a}^\dag + \hat{a} \right), \qquad \hat{p}(0) = i \sqrt{\frac{\hbar}{2}} \left( \hat{a}^\dag - \hat{a} \right). \eeq
Evolving operators $\hat{q}$ and $\hat{p}$ with the Hamiltonian~\eqref{eq:harmonic-H-ladder}, we find their explicit form at an arbitrary moment:
\beq \label{eq:harmonic-qp-t}
\hat{q}(t) = \sqrt{\frac{\hbar}{2}} \left( \hat{a}^\dag e^{i t} + \hat{a} e^{-i t} \right), \qquad \hat{p}(t) = i \sqrt{\frac{\hbar}{2}} \left( \hat{a}^\dag e^{i t} - \hat{a} e^{-i t} \right). \eeq
Now, let us substitute operators~\eqref{eq:harmonic-qp-0} and~\eqref{eq:harmonic-qp-t} into the definition of the ROTOC~\eqref{eq:ROTOC-def}:
\beq \label{eq:harmonic-ROTOC} \begin{aligned}
C_n(t) &= \left\langle \left[ -\frac{1}{2 \hbar^2} \left( \big[ \hat{q}(t), \hat{q}(0) \big]^2 + \big[ \hat{q}(t), \hat{p}(0) \big]^2 + \big[ \hat{p}(t), \hat{q}(0) \big]^2 + \big[ \hat{p}(t), \hat{p}(0) \big]^2\right) \right]^n \right\rangle \\ &= \left\langle \left[ \frac{1}{2} \left( \sin^2(t) + \cos^2(t) + \cos^2(t) + \sin^2(t) \right) \right]^n \right\rangle = 1.
\end{aligned} \eeq
We do not even need to specify the averaging because the commutators $\left[ z_i(t), z_j(0) \right]$ are c-numbers.

Finally, we calculate the LOTOC~\eqref{eq:LOTOC-def} and the refined quantum LE~\eqref{eq:true-qLE-def}:
\beq \label{eq:harmonic-qLE}
L(t) = \lim_{n \to 0} \frac{\pd C_n(t)}{\pd n} = 0, \qquad \bar{\kappa}_q = 0. \eeq
Therefore, the quantum harmonic oscillator is not chaotic, which perfectly agrees with the integrability of classical harmonic oscillator.

We can also formally quantize the inverted harmonic oscillator~\cite{Bhattacharyya:2020, Wang:2022, Morita}:
\beq \label{eq:harmonic-inv-H}
\hat{H}^\mathrm{inv} = \frac{1}{2} \hat{p}^2 - \frac{1}{2} \hat{q}^2, \eeq
by substituting the decomposition of the position and momentum operators~\eqref{eq:harmonic-qp-0} into the Hamiltonian~\eqref{eq:harmonic-inv-H}. Evolving these operators with such a formally derived Hamiltonian, we obtain analogs of Eqs.~\eqref{eq:harmonic-qp-t}:
\beq \label{eq:harmonic-inv-qp}
\begin{gathered}
\hat{q}^\mathrm{inv}(t) = \sqrt{\frac{\hbar}{2}} \left[ \hat{a}^\dag \left( \cosh(t) + i \sinh(t) \right) + \hat{a} \left( \cosh(t) - i \sinh(t) \right) \right], \\
\hat{p}^\mathrm{inv}(t) = \sqrt{\frac{\hbar}{2}} \left[ \hat{a}^\dag \left( \sinh(t) + i \cosh(t) \right) + \hat{a} \left( \sinh(t) - i \cosh(t) \right) \right],
\end{gathered} \eeq
and the ROTOC~\eqref{eq:harmonic-ROTOC}:
\beq \label{eq:harmonic-inv-ROTOC}
C_n^\mathrm{inv} = \left\langle \left[ \sinh^2(t) + \cosh^2(t) \right]^n \right\rangle \stackrel{!}{=} \left(e^{2t} + e^{-2t}\right)^n/2^n. \eeq
So we estimate the LOTOC:
\beq \label{eq:harmonic-inv-LOTOC}
L(t) \stackrel{!}{=} \log\left(e^{2t} + e^{-2t} \right) - \log 2 = 2t + o(t), \quad \text{where} \quad o(t) = \log\left( 1 + e^{-4t} \right) -\log 2, \eeq
and the refined quantum LE:
\beq \label{eq:harmonic-inv-qLE}
\bar{\kappa}_q \stackrel{!}{=} 1. \eeq
This estimate also formally coincides with the average classical LE~\eqref{eq:cl-LE-3}:
\beq \label{eq:harmonic-inv-cl}
\begin{cases} \dot{q} = p, \\ \dot{p} = q, \end{cases} \Longrightarrow \quad \begin{cases} q(t) = q_0 \cosh(t) + p_0 \sinh(t), \\ p(t) = q_0 \sinh(t) + p_0 \cosh(t), \end{cases} \Longrightarrow \quad \bar{\kappa}_{cl} \stackrel{!}{=} 1. \eeq

However, we emphasize that the spectrum of the inverted oscillator is unbounded from below, so the averaging in both classical and quantum cases is ill-defined; we underline this by writing ``$\stackrel{!}{=}$'' instead of ``$=$'' in all equalities that involve averaging. Hence, the status of Eqs.~\eqref{eq:harmonic-inv-ROTOC}--\eqref{eq:harmonic-inv-cl} is uncertain, although all expressions under the averages~\eqref{eq:harmonic-inv-ROTOC}--\eqref{eq:harmonic-inv-cl} do not depend on the averaging and the correspondence between the classical and quantum LEs formally holds.

To resolve the problem with averaging in the model~\eqref{eq:harmonic-inv-H}, we need to constrain the minimum energy; e.g., we can impose hard-wall boundary conditions $\psi(\pm L) = 0$ at a finite distance $L$ or add higher-order corrections to the potential energy, $V(q) \to -\frac{m^2}{2} q^2 + \frac{\lambda}{4} q^4$. In both cases, the exponential divergence of classical trajectories is observed near the saddle point, but slows down eventually due to integrability of the system\footnote{One-dimensional Hamiltonian systems are automatically integrable due to Liouville's theorem.}, so definitions~\eqref{eq:cl-LE-1} and~\eqref{eq:cl-LE-2} give zero classical LEs everywhere except the saddle point. Therefore, we expect $\bar{\kappa}_{cl} = 0$ and $\bar{\kappa}_q = 0$. We will discuss such a situation in the following subsection.

\subsection{Lipkin-Meshkov-Glick model}
\label{sec:LMG}

As a convenient example of an integrable model with an isolated saddle point, we consider a special instance of the Lipkin-Meshkov-Glick (LMG) model~\cite{LMG-1, LMG-2, LMG-3}, which is very similar to a particle in a double-well potential:
\beq \label{eq:LMG-H}
H = x + 2 z^2. \eeq
Here, $x,y,z$ form a classical $SU(2)$ spin that lives on the unit sphere, $x^2 + y^2 + z^2 = 1$, and has the following Poisson bracket: $\{ x_m, x_n \} = \epsilon_{m n k} x_k$. The equations of motion are as follows:
\beq \label{eq:LMG-eom}
\dot{x}_m = \{ H, x_m \} \Longleftrightarrow \begin{cases} \dot{x} = 4 y z, \\ \dot{y} = z - 4 x z, \\ \dot{z} = -y. \end{cases} \eeq
Linearizing Eqs.~\eqref{eq:LMG-eom}, we find four fixed points:
\beq \label{eq:LMG-fixed}
(x,y,z) = (\pm 1, 0, 0) \qquad \text{and} \qquad (x,y,z) = \left( 1/4, \, 0, \, \pm \sqrt{15}/4 \right). \eeq
However, only the point $(x,y,z) = (1,0,0)$ is unstable (i.e., a saddle point):
\beq \label{eq:LMG-saddle}
\begin{cases} \dot{\delta x} \approx 0, \\ \dot{\delta y} \approx -3 \delta z, \\ \dot{\delta z} \approx - \delta y \end{cases} \Longrightarrow \quad \begin{cases} \delta x(t) \approx \delta x_0, \\ \kappa_s \delta z(t) - \delta y(t) \approx \left( \kappa_s \delta z_0 - \delta y_0 \right) e^{\kappa_s t}, \\ \kappa_s \delta z(t) + \delta y(t) \approx \left( \kappa_s \delta z_0 + \delta y_0 \right) e^{-\kappa_s t}, \end{cases}\eeq
where we expand the coordinates near the point, $(x,y,z) \approx (1,0,0) + (\delta x, \delta y, \delta z)$, and introduce $\kappa_s = \sqrt{3}$. Similarly to a particle in a double-well potential, Eqs.~\eqref{eq:double-well-H}--\eqref{eq:double-well-sensitivity}, the exponential growth~\eqref{eq:LMG-saddle} is observed only in a narrow strip near the saddle point:
\beq \mathcal{S}_{t_*} = \left\{ | \kappa_s \delta z_0 - \delta y_0 | < \delta e^{-\kappa_s t_*}, \; | \kappa_s \delta z_0 + \delta y_0 | < \delta, \; \big(1 + \delta x_0\big)^2 + \delta y_0^2 + \delta z_0^2 = 1 \right\}, \eeq
where $\delta \ll 1$, and only until $t < t_*$. Therefore, in the semiclassical limit, the ROTOCs are estimated similarly to a simple OTOC~\eqref{eq:double-well-OTOC}:
\beq \label{eq:LMG-ROTOC}
C_n(t) \sim \int_{\mathcal{S}_t} d\mathbf{z}_0 \left[ \frac{1}{2} \sum_{i,j} \Phi_{ji}^*(t; \mathbf{z}_0) \Phi_{ij}(t; \mathbf{z}_0) \right]^n \! e^{-\beta H(\mathbf{z}_0)} \sim \delta^2 e^{-t} \times \left[ e^{2 \kappa_s t} \right]^n \sim \delta^2 e^{(2n - 1) \kappa_s t}. \eeq
We emphasize that the ROTOCs grow exponentially only when $n > n_* = 1/2$. More precisely, we expect that the replica LE $\kappa_n$, which is extracted from the exponential growth of the ROTOC up to the Ehrenfest time, $C_n(t) \sim e^{2 \kappa_n t}$, is an analytic function of $n$ that interpolates between $\kappa_n = (n - 1/2) \kappa_s$ for integer $n > n_*$ and $\kappa_n = 0$ for $n < n_*$. Hence, when we analytically continue~\eqref{eq:LMG-ROTOC} to real $n$ and take the limit $n \to 0$, the replica trick~\eqref{eq:replica-trick} gives \textit{zero} refined quantum LE, although ROTOCs grow exponentially for every single positive integer $n$:
\beq \label{eq:LMG-qLE}
\bar{\kappa}_q = \lim_{n \to 0} \frac{\pd \kappa_n}{\pd n} = \lim_{n \to 0} \frac{\pd \kappa_n}{\pd n} \Big|_{n < n_*} = 0. \eeq

\begin{figure}[t]
    \centering
    \includegraphics[width=\linewidth]{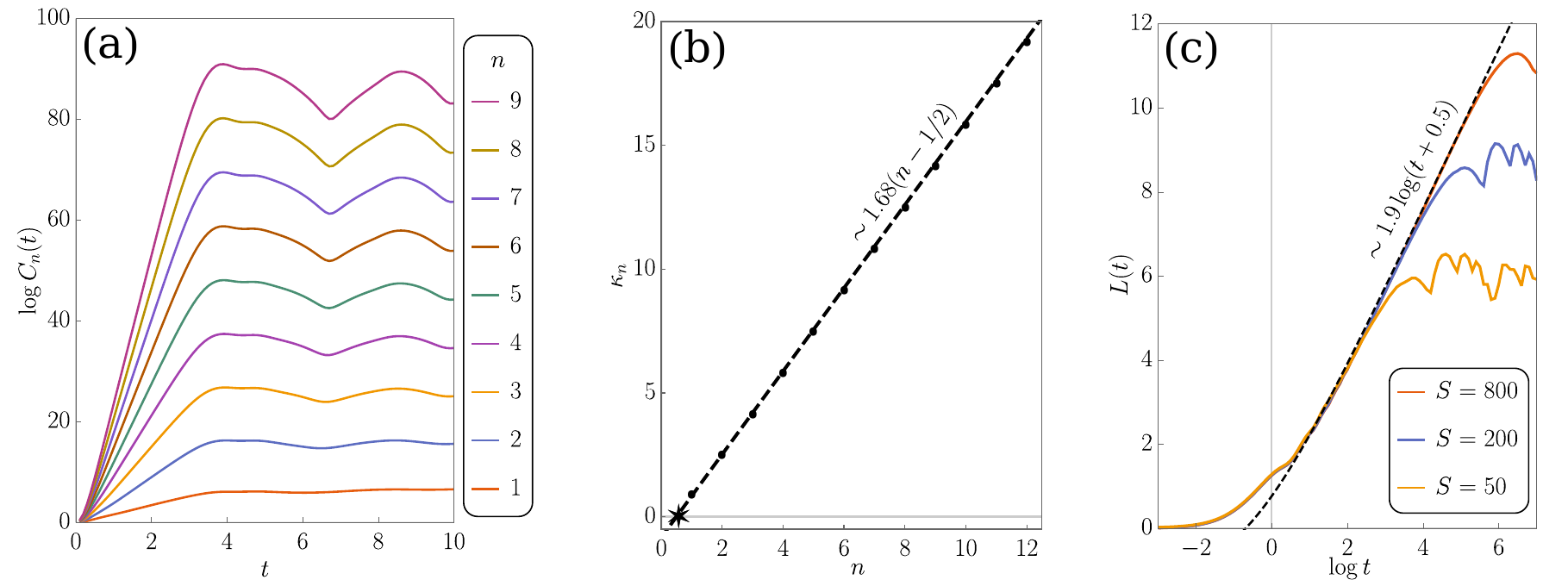}
    \caption{(a) First nine ROTOCs in the infinite-temperature quantum LMG model with $S = 200$. (b) Replica LEs $\kappa_n$ extracted from the infinite-temperature ROTOCs, $C_n(t) \sim e^{2 \kappa_n t}$ as $1 \ll t \ll t_E$. The asterisk marks the intersection of the $x$-axis and the dashed line, which extrapolates the numerical points. (c) Infinite-temperature LOTOCs in the logarithmic timescale. Note that the ROTOCs saturate at the time $t_E^\mathrm{chaotic} \sim \log S \sim \log(1/\hbar)$, which is determined by the saddle point, and the LOTOCs saturate at the time $t_E^\mathrm{integrable} \sim S \sim 1/\hbar$, which reflects the integrability of the LMG model as a whole.}
    \label{fig:LMG}
\end{figure}
The numerical calculations in the quantum LMG model (Fig.~\ref{fig:LMG}) corroborate this qualitative reasoning. To perform these calculations, we quantize the model~\eqref{eq:LMG-H} by replacing the components of the classical spin with corresponding rescaled operators from a $SU(2)$ representation with a large total spin $S$: $(\hat{x}, \hat{y}, \hat{z}) = \left( \hat{S}_x/S, \hat{S}_y/S, \hat{S}_z/S \right)$, so $\hat{x}^2 + \hat{y}^2 + \hat{z}^2 = 1 + 1/S \approx 1$ as $S \to \infty$ and $[\hat{x}_m, \hat{x}_n ] = i \hbar \epsilon_{m n k} \hat{x}_k$ with $\hbar = 1/S$. Then we write down the quantized Hamiltonian~\eqref{eq:LMG-H} in the basis of eigenfunctions of $\hat{S}_z$, numerically solve the stationary Schr{\"o}dinger equation, estimate the evolution operator, and explicitly calculate the LOTOC~\eqref{eq:LOTOC-def} and the ROTOCs~\eqref{eq:ROTOC-def}. For convenience, we work in $(x,y,z)$ coordinates. Besides, we consider the infinite-temperature limit, where the behavior of the ROTOCs and the LOTOC is most pronounced. As a result, we confirm the exponential growth of the ROTOCs~\eqref{eq:LMG-ROTOC}, see Fig.~\ref{fig:LMG}(a,b). We also establish the logarithmic growth of the LOTOC up to the Ehrenfest time $t_E \sim 1/\hbar$, see Fig.~\ref{fig:LMG}(c), which yields a \textit{zero} refined quantum LE~\eqref{eq:LMG-qLE}:
\beq \label{eq:LMG-numerics}
L(t) \approx 2 \bar{\kappa}_q t + o(t), \quad \text{where} \quad \bar{\kappa}_q = 0 \quad \text{and} \quad o(t) \approx 1.9 \log(t + 0.5). \eeq
Since the classical LMG model is integrable, this result again establishes the correspondence between $\bar{\kappa}_{cl} = 0$ and $\bar{\kappa}_q = 0$.

\section{Quantum chaotic map}
\label{sec:map}

Another famous testing ground for the correspondence between classical and quantum chaos is the Arnold cat map~\cite{Arnold} and its quantum counterpart~\cite{Esposti, Hannay, Keating, Andries, Neshveyev, Garcia-Mata:2018, Moudgalya, Chen:2018}. In this section, we calculate the ROTOCs and the refined quantum LE for this model.

The classical Arnold cat map acts on the unit torus $\mathbb{T}^2 = \mathbb{R}^2/\mathbb{Z}^2$:
\beq \label{eq:cat-classical}
\bem q \\ p \eem \to \bem 2 & 1 \\ 1 & 1 \eem \bem q \\ p \eem \mod \; 1, \eeq
preserves the area, and has a positive largest LE for all initial conditions: $\bar{\kappa}_{cl} = \log\!\left[ \left( 3 + \sqrt{5} \right) / 2 \right]$. 

To quantize the map~\eqref{eq:cat-classical}, we first notice that the wave functions on a unit torus must be periodic both in position $q$ and momentum $p$:
\beq \label{eq:torus-wave-function}
\psi(q+1) = \psi(q) \qquad \text{and} \qquad \widetilde{\psi}(p+1) = \widetilde{\psi}(p), \eeq
where the Fourier transform is defined as follows:
\beq \widetilde{\psi}(p) = \frac{1}{2 \pi \hbar} \int_{-\infty}^\infty \psi(q) e^{-i q p / \hbar} dq. \eeq
The periodicity~\eqref{eq:torus-wave-function} readily implies that the Hilbert space has a finite integer dimension $N > 0$, the Planck's constant is $\hbar = 1 / 2 \pi N$, and the semiclassical limit corresponds to $N \to \infty$. Hence, we can define a discrete position basis $\left\{ | m \rangle = \delta(q - m/N) \right\}$, where $m = 0, \,\cdots, N-1$, decompose the wave functions~\eqref{eq:torus-wave-function}:
\beq \label{eq:torus-wave-function-decomposition}
\psi(q) = \sum_{m = 0}^{N-1} \psi_m | m \rangle, \eeq
and write down the unitary operator $\hat{U}$ that defines a quantum analog of the map~\eqref{eq:cat-classical} \cite{Esposti, Hannay, Keating}:
\beq \label{eq:cat-quantum}
\hat{U} = \sum_{m, n = 0}^{N-1} \frac{1}{\sqrt{N}} \exp\left[ \frac{2 i \pi}{N} \left( m^2 - m n + \frac{n^2}{2} \right) \right] | n \rangle \langle m |. \eeq
Then, the time evolution of the wave functions~\eqref{eq:torus-wave-function} is given in discrete steps, $t = 1, 2, \cdots$, by $\hat{U}^t$.

More precisely, the canonical quantization on torus is obtained upon classification of all irreducible representations of the discrete Heisenberg group defined by the following relations:
\beq \label{eq:discrete-Heisenberg-group}
\begin{gathered}
T^\dag(\mathbf{z}) = T(-\mathbf{z}), \\
T(\mathbf{z}) T(\mathbf{z}') = e^{(i \pi / N) \omega(\mathbf{z}, \mathbf{z}')} T(\mathbf{z} + \mathbf{z}'),
\end{gathered} \eeq
where we introduce short notations for the phase space coordinate, $\mathbf{z} = (q,p)$, and the symplectic product, $\omega(\mathbf{z}, \mathbf{z}') = q p' - q' p$. Such representations can be built by specifying the action of the discrete Heisenberg algebra generators $\hat{t}_1$ and $\hat{t}_2$ on the wave functions~\eqref{eq:torus-wave-function-decomposition}\footnote{One can similarly construct representations that do not satisfy periodic boundary conditions.}:
\beq \label{eq:discrete-Heisenberg-generators}
\hat{t}_1 | m \rangle = e^{2 i \pi m/N} | m \rangle, \qquad \hat{t}_2 | m \rangle = | m + 1 \rangle. \eeq
Then we easily construct the representation for an arbitrary group element $T(\mathbf{z})$, which is nothing but the Weyl translation operator:
\beq \label{eq:Weyl-operators}
\hat{T}(\mathbf{z}) = e^{i \pi q p / N} \, \hat{t}_2^p \, \hat{t}_1^q = e^{i \pi q p / N} \sum_{m = 0}^{N-1} e^{2 i \pi q m / N} | m + p \rangle \langle m |. \eeq
In fact, it is straightforward to see that the generators $\hat{t}_1 = \hat{T}(1,0)$ and $\hat{t}_2 = \hat{T}(0,1)$ correspond to the exponents of the conventional position and momentum operators~\cite{Esposti, Santhanam}:
\beq \label{eq:torus-qp}
\hat{q} \psi(q) = q \psi(q), \qquad \hat{p} \psi(q) = -i \hbar \psi'(q), \eeq
so the commutation relation $\left[ \hat{q}, \hat{p} \right] = i \hbar$ is equivalent to the relation $\hat{t}_1 \hat{t}_2 = e^{2 i \pi / N} \hat{t}_2 \hat{t}_1$, which generates the group action~\eqref{eq:discrete-Heisenberg-group}. Furthermore, in the semiclassical limit, we can decompose these exponents and approximate the position and momentum operators as follows:
\beq \label{eq:torus-qp-approximation}
\hat{q} = \frac{\hat{t}_1 - \hat{t}_1^\dag}{2i} \qquad \text{and} \qquad \hat{p} = \frac{\hat{t}_2 - \hat{t}_2^\dag}{2i} \qquad \text{as} \qquad \hbar \to 0. \eeq
Besides, we note that the translation operators~\eqref{eq:Weyl-operators} transform ``classically'' (by construction) during the discrete time evolution~\eqref{eq:cat-quantum}:
\beq \label{eq:Weyl-evolution}
\big(\hat{U}^\dag\big)^t \, \hat{T}(\mathbf{z}) \, \hat{U}^t = \hat{T}(\mathbf{M}^t \mathbf{z}), \eeq
where matrix $\mathbf{M} = \bem 2 & 1 \\ 1 & 1 \eem$ generates the classical evolution~\eqref{eq:cat-classical}, i.e., $\mathbf{z}(t+1) = \mathbf{M} \mathbf{z}(t)$.

To calculate the ROTOCs for the quantized cat map~\eqref{eq:cat-classical}, we also need the commutation relation on operators $\hat{F}(\mathbf{z}) = \frac{\hat{T}(\mathbf{z}) - \hat{T}^\dag(\mathbf{z})}{2 i}$, which follows from Eq.~\eqref{eq:discrete-Heisenberg-group}:
\beq \label{eq:cat-F}
\left[ \hat{F}(\mathbf{z}), \hat{F}(\mathbf{z}') \right] = \frac{1}{2 i} \sin\left[ \frac{\pi \omega(\mathbf{z}, \mathbf{z}')}{N} \right] \left[ \hat{T}(\mathbf{z} + \mathbf{z}') + \hat{T}(\mathbf{z} - \mathbf{z}') + \hat{T}(-\mathbf{z} + \mathbf{z}') + \hat{T}(-\mathbf{z} - \mathbf{z}') \right]. \eeq
Representing operators~\eqref{eq:torus-qp-approximation} as $\hat{q} = \hat{F}(\mathbf{q})$ and $\hat{p} = \hat{F}(\mathbf{p})$, where we denote $\mathbf{q} = (1,0)$ and $\mathbf{p} = (0,1)$ for shortness, and employing the relations~\eqref{eq:Weyl-evolution},~\eqref{eq:cat-F}, we obtain the ROTOC:
\beq \label{eq:cat-ROTOC} \begin{aligned}
C_n(t) &= \left\langle \left[ -\frac{1}{2 \hbar^2} \left( \big[ \hat{q}(t), \hat{q} \big]^2 + \big[ \hat{q}(t), \hat{p} \big]^2 + \big[ \hat{p}(t), \hat{q} \big]^2 + \big[ \hat{p}(t), \hat{p} \big]^2\right) \right]^n \right\rangle \\ &= \left\langle \left[ \frac{1}{2 \hbar^2} \left( 2 \sin^2 \frac{\pi \omega(\mathbf{M}^t \mathbf{q}, \mathbf{q})}{N} + \sin^2 \frac{\pi \omega(\mathbf{M}^t \mathbf{q}, \mathbf{p})}{N} + \sin^2 \frac{\pi \omega(\mathbf{M}^t \mathbf{p}, \mathbf{q})}{N} \right) \left( \hat{T}(0,0) + \cdots \right) \right]^n \right\rangle.
\end{aligned} \eeq
Here, the angle brackets denote the averaging over an infinite-temperature thermal ensemble, $\langle \,\cdots \rangle = \tr( \,\cdots ) / N$, and the ellipsis in the last line hides all the operators different from the identity operator $\hat{T}(0,0)$, e.g., $\hat{T}(2 \mathbf{M}^t \mathbf{q} - 2 \mathbf{p})$ or $\hat{T}(2 \mathbf{q})$.

\begin{figure}[t]
    \centering
    \includegraphics[width=\linewidth]{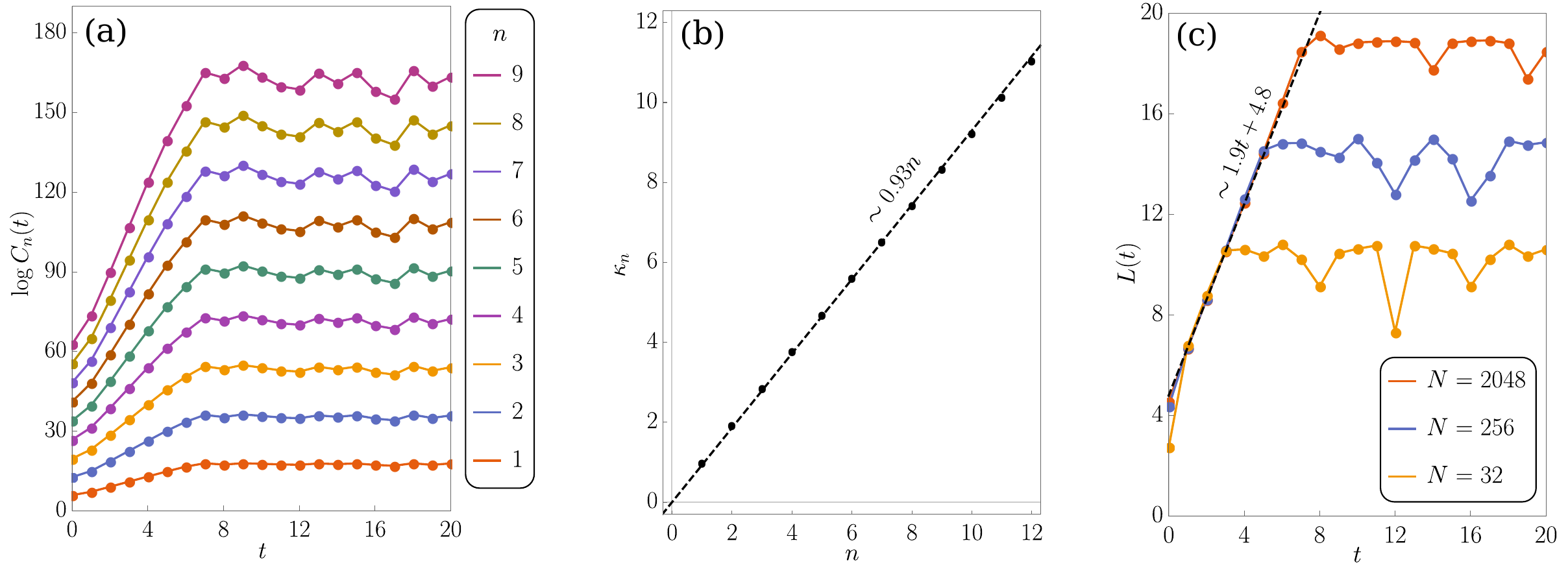}
    \caption{(a) Infinite-temperature ROTOCs in the quantized version of the cat map~\eqref{eq:cat-classical} with the Hilbert space dimension $N = 1024$. (b) Replica LEs $\kappa_n$ extracted from the infinite-temperature ROTOCs using~\eqref{eq:cat-ROTOC-final}. Unlike the integrable LMG model, Fig.~\ref{fig:LMG}(b), in the chaotic quantum map, replica LEs remain positive even as $n \to 0$, which implies a nonzero refined quantum LE. (c) Infinite-temperature LOTOCs. We emphasize that \textit{both} ROTOCs and LOTOCs saturate at the chaotic Ehrenfest time $t_E^\mathrm{chaotic} \sim \log(N) \sim \log(1/\hbar)$.}
    \label{fig:cat}
\end{figure}
To estimate the average~\eqref{eq:cat-ROTOC}, we make two key observations. First, all four elements of matrix $\mathbf{M}^t$ exponentially grow until they become of order $N$ --- we remind that due to periodicity of wave functions~\eqref{eq:torus-wave-function-decomposition}, this matrix acts on a finite-dimensional phase space, $\mathbf{z} \sim \mathbf{z} \mod N$. Hence, all symplectic products in~\eqref{eq:cat-ROTOC} are estimated as the power of the largest eigenvalue of $\mathbf{M}$, i.e., $\omega(\mathbf{M}^t \mathbf{q}, \mathbf{q}) \sim \left[ (3 + \sqrt{5})/2 \right]^t = e^{\bar{\kappa}_{cl} t}$ for $1 \ll t \ll \log(N)$ and so on. Second, all translation operators except the identical one are traceless, $\tr \big[ \hat{T}(\mathbf{z}) \big] = 0$ if $\mathbf{z} \neq (0,0)$. Moreover, the prefactors of operators hidden in ellipsis in~\eqref{eq:cat-ROTOC} only very slightly depend on $t$ as $t \ll \log(N)$. Therefore, the trace of the operator sum in~\eqref{eq:cat-ROTOC} gives a mere combinatorial factor\footnote{To determine the combinatorial factor $A_n$, one needs to calculate the total weight of all possible paths that start and end at the same point on a toric $N \times N$ grid, have length up to $n$, and are generated by weighted steps on vectors $\mathbf{q}$, $\mathbf{p}$, $\mathbf{M}^t \mathbf{q}$, $\mathbf{M}^t \mathbf{p}$ (approximate weight $1/2$) and $\mathbf{M}^t \mathbf{q} \pm \mathbf{q}$, $\mathbf{M}^t \mathbf{q} \pm \mathbf{p}$, $\mathbf{M}^t \mathbf{p} \pm \mathbf{q}$, $\mathbf{M}^t \mathbf{p} \pm \mathbf{p}$ (weight $1/4$).}, $\tr \big[ \hat{T}(0,0) + \cdots \big]^n \approx A_n$ as $t \ll \log(N)$. Thus we estimate the ROTOC:
\beq \label{eq:cat-ROTOC-final}
C_n(t) \approx A_n \left( 2 \pi^4 \right)^n e^{2 n \bar{\kappa}_{cl} t} \quad \text{as} \quad 1 \ll t \ll t_E, \eeq
the LOTOC:
\beq \label{eq:cat-LOTOC}
L(t) \approx 2 \bar{\kappa}_{cl} t + \const \quad \text{as} \quad 1 \ll t \ll t_E, \eeq
and the refined quantum LE:
\beq \label{eq:cat-qLE}
\bar{\kappa}_q \approx \bar{\kappa}_{cl} \approx 0.96. \eeq
Here, the Ehrenfest time scales logarithmically with the Planck's constant, $t_E \sim \log(1/\hbar)$, due to chaoticity of the classical map~\eqref{eq:cat-classical} and its quantum counterpart. Finally, the numerical calculations (Fig.~\ref{fig:cat}) confirm the estimates~\eqref{eq:cat-ROTOC-final}--\eqref{eq:cat-qLE}.

\section{Diagram calculations}
\label{sec:diagrams}

Diagrammatic expansion is one of the most powerful tools for calculating correlation functions, including OTOCs. There are two key assumptions that underlie any diagrammatics. First is the possibility of perturbative expansion, which means that correlation functions in an interacting theory can be expressed in terms of correlation functions in a free theory. Second is the Wick's theorem, which relates a product of operators ordered along a time contour --- generally speaking, an arbitrary time contour on a complex plane --- to a sum of all possible normal-ordered contractions. If the initial state of a system coincides with that of a free theory (e.g, if all coupling constants are adiabatically switched off at the past infinity), then the Wick's theorem implies that an arbitrary correlator can be decomposed into a sum of all possible products of the corresponding two-point functions.

In the simplest case of time-ordered correlators, these two assumptions lead to the Feynman's diagram technique~\cite{Peskin}. In that case, the time ordering is performed along a simple line that stretches from the past to the future infinity along a real axis. If we want to calculate arbitrary-ordered two-point functions or related quantities, we need to replace this contour with the Keldysh one, which contains parts with both forward and backward evolution in time. Thus we obtain the Schwinger-Keldysh diagram technique~\cite{Keldysh, Arseev, Berges}. Finally, if we want to cover all arbitrary-ordered four-point functions, we have no choice but to generalize the construction of the Schwinger-Keldysh technique to a two-fold Keldysh contour, which comprises two separate parts with forward and backward evolution in time. This construction allows one to calculate the OTOCs using the standard approaches of the Feynman and Schwinger-Keldysh techniques, and is sometimes referred to as the augmented Schwinger-Keldysh technique~\cite{Aleiner, Haehl}.

In this section, we continue the sequence of diagrammatic generalizations to $2n$-fold contours that allow us to calculate arbitrary-ordered $4n$-point functions, including the ROTOC $C_n(t)$. More precisely, we consider a regularized version of~\eqref{eq:ROTOC-def}:
\beq \label{eq:ROTOC-reg}
C_n^\mathrm{reg}(t) = \frac{1}{\tr \big( \hat{y}^{n+1} \big)} \tr\!\left[ \hat{y} \, \bigg( \frac{1}{N \hbar^2} \sum_{i,j} \hat{y}^{1/2} \left[ \hat{z}_i(t), \hat{z}_j(0) \right]^\dag \hat{y}^{1/2} \left[ \hat{z}_i(t), \hat{z}_j(0) \right] \bigg)^n \right], \eeq
Here, $\hat{y} = e^{- \beta \hat{H}}$, $\beta = 1 / T$ is the inverse temperature, so that $\hat{\rho} = \hat{y} / \tr \big( \hat{y} \big)$ is a thermal density matrix. The regularization of~\eqref{eq:ROTOC-def} is required because many operators there are taken at the same real times and achieved in~\eqref{eq:ROTOC-reg} by pushing operators apart in imaginary time. In quantum mechanics, this operation is usually idle, but it saves us from potential infrared divergences and is especially important for calculating ROTOCs in quantum field theories.

On the one hand, the regularized ROTOC and LOTOC are related similarly to~\eqref{eq:replica-trick}:
\beq \label{eq:LOTOC-reg}
L^\mathrm{reg}(t) = \lim_{n \to 0} \frac{\pd C_n^\mathrm{reg}(t)}{\pd n} = \tr\!\left[ \hat{\rho} \log\!\bigg( \frac{1}{N \hbar^2} \sum_{i,j} \hat{\rho}^{1/2} \left[ \hat{z}_i(t), \hat{z}_j(0) \right]^\dag \hat{\rho}^{1/2} \left[ \hat{z}_i(t), \hat{z}_j(0) \right] \bigg) \right] - \tr \left[ \hat{\rho} \log \hat{\rho} \right], \eeq
where the time-independent contribution is just the entropy of the initial quantum state. Furthermore, the regularization~\eqref{eq:LOTOC-reg} corresponds to shifting $\hat{z}_i(t) \to \hat{z}_i(t + i \beta / 4)$ and $\hat{z}_j(0) \to \hat{z}_j(i \beta / 4)$ in both commutators in~\eqref{eq:LOTOC-def}. Such a shift does not affect the growth of the correlator, which depends only on the difference of these two times. Hence, the regularized versions of ROTOCs and LOTOC distinguish between quantum integrable and chaotic models just as good as their unregularized versions~\eqref{eq:LOTOC-def},~\eqref{eq:ROTOC-def}.

\begin{figure}[t]
    \center{\includegraphics[width=0.75\linewidth]{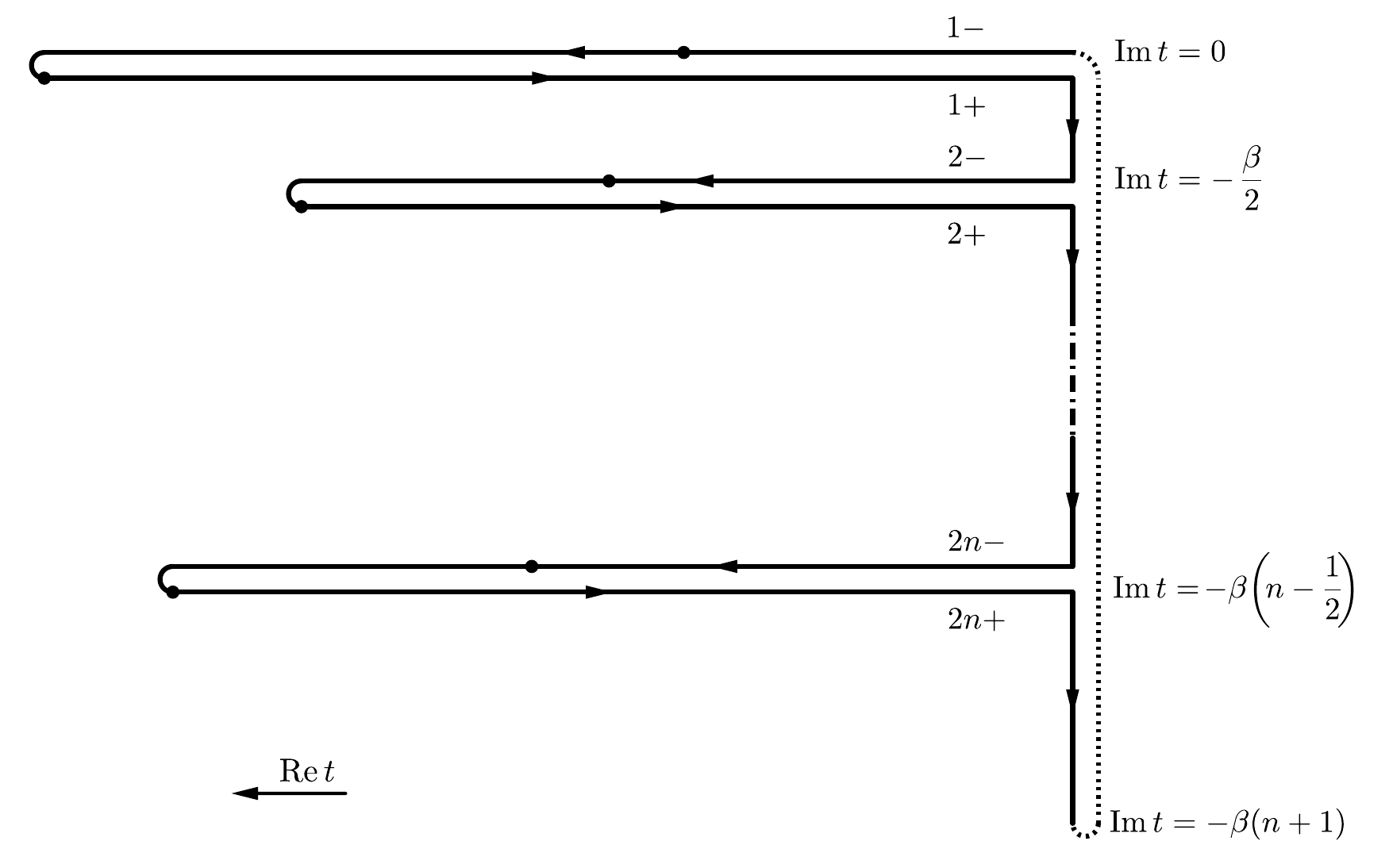}}
    \caption{The $2n$-fold Keldysh contour $\mathcal{C}$ that allows us to rewrite an arbitrary-ordered $4n$-point correlation function in a time-ordered from~\eqref{eq:technique-correlator-time-ordered}. Horizontal lines denote the evolution in real time, vertical lines denote the evolution in imaginary time, and the dotted line denotes the identification of points $t = t_0$ and $t = t_0 - i \beta (n+1)$. Black dots mark the insertions of operators into the correlation function. The arrow of real time is directed to the left.}
    \label{fig:contour}
\end{figure}

On the other hand, the regularized correlation function~\eqref{eq:ROTOC-reg} can be treated as a sum of $4n$-point correlation functions defined on a Keldysh contour with $2n$ real folds and periodic imaginary part, which corresponds to the effective temperature $T_n = T / (n+1)$, see Fig.~\ref{fig:contour}. In the following subsections, we introduce a general approach to calculating such correlation functions and discuss several particular examples. For brevity, we set $\hbar = 1$ in this section.

\subsection{Schwinger-Keldysh technique on a multi-fold Keldysh contour}
\label{sec:basics}

Our main goal is to calculate the correlation functions of the following type:
\beq \label{eq:technique-general-correlator}
\tilde{C}(t_1, \cdots, t_{4n}) = \frac{1}{\tr(\hat{y}^{n+1})} \tr\!\left[ \hat{y} \, \hat{y}^{1/2} \hat{O}(t_{4n}) \hat{O}(t_{4n-1}) \cdots \hat{y}^{1/2} \hat{O}(t_4) \hat{O}(t_3) \, \hat{y}^{1/2} \hat{O}(t_2) \hat{O}(t_1) \right]. \eeq
Here, $\hat{y} = e^{-\beta \hat{H}}$, $\hat{O}(t) = \hat{U}^\dag(t, t_0) \hat{O}(t_0) \hat{U}(t, t_0)$ are some operators in the Heisenberg picture, $\hat{U}(t, t_0) = \mathrm{\hat{T}exp}\left[ -i \int_{t_0}^t \hat{H}(t') dt' \right]$ is the evolution operator in a model with the Hamiltonian $\hat{H}$, and $t_0 \to -\infty$. Note that $\hat{y} = e^{-\beta \hat{H}} = \hat{U}(t_0 - i \beta, t_0)$ can be considered as an evolution operator in the imaginary time if $\hat{H}$ does not depend on time in the asymptotic past. For illustrative purposes, in this subsection, we consider a $(0+1)$-dimensional $\lambda \phi^4$ model (quartic oscillator) and take $\hat{O} = \hat{x}$, although generalizations to arbitrary models and correlators are straightforward:
\beq \label{eq:H-phi-4}
\hat{H} = \frac{1}{2} \hat{p}^2 + \frac{1}{2} m^2 \hat{x}^2 + \frac{\lambda}{4} \hat{x}^4. \eeq

First of all, we rewrite the correlator~\eqref{eq:technique-general-correlator} in the interaction picture:
\beq \label{eq:technique-correlator-interaction-picture}
\tilde{C}(t_1, \cdots, t_{4n}) = \frac{1}{\tr(\hat{y}^{n+1})} \tr\left[ \hat{y} \, \hat{y}^{1/2} \hat{U}_I^\dag (t_{4n}, t_0) \hat{x}_0(t_{4n}) \hat{U}_I (t_{4n}, t_{4n-1}) \hat{x}_0(t_{4n-1}) \hat{U}_I (t_{4n-1}, t_0) \cdots \right], \eeq
where $\hat{x}_0(t) = \hat{U}_0^\dag(t, t_0) \hat{x}(t_0) \hat{U}_0(t, t_0)$, $\hat{U}_0(t, t_0) = \mathrm{\hat{T}exp}\left[ - i \int_{t_0}^t \hat{H}_0(t') dt' \right]$ is the evolution operator in the free theory with Hamiltonian $\hat{H}_0 = \frac{1}{2} \hat{p}^2 + \frac{1}{2} m^2 \hat{x}^2$, and $\hat{U}_I(t, t_0) = \mathrm{\hat{T}exp}\left[ - i \int_{t_0}^t \hat{H}_\mathrm{int}(t') dt' \right]$ with $\hat{H}_\mathrm{int} = \frac{\lambda}{4} \hat{x}^4$. Then we introduce the $2n$-fold Keldysh contour $\mathcal{C}$ (Fig.~\ref{fig:contour}), assign all operators to different branches of this contour, and rewrite the correlator~\eqref{eq:technique-correlator-interaction-picture} in a time-ordered form:
\beq \label{eq:technique-correlator-time-ordered}
\begin{aligned}
\tilde{C}(t_1, \cdots, t_{4n}) &= \tr \, \mT_\mathcal{C} \left[ \hat{\rho}_{n+1} \, \hat{x}_{1,-}(t_1) \hat{x}_{1,+}(t_2) \cdots \hat{x}_{2n,-}(t_{4n-1}) \hat{x}_{2n,+}(t_{4n}) \exp\left( -i \int_\mathcal{C} \hat{H}_\mathrm{int}(t') dt' \right) \right] \\ &\equiv \left\langle \hat{x}_{1,-}(t_1) \hat{x}_{1,+}(t_2) \cdots \hat{x}_{2n,-}(t_{4n-1}) \hat{x}_{2n,+}(t_{4n}) \right\rangle,
\end{aligned} \eeq
where we suppress the index 0 for brevity, introduce the effective density matrix $\hat{\rho}_{n+1} = \hat{y}^{n+1} / \tr\big( \hat{y}^{n+1} \big)$, and denote the ordering along the contour $\mathcal{C}$ as $\mT_\mathcal{C}$. Essentially, the ``plus'' (``minus'') horizontal branches correspond to the forward (backward) evolution in real time, and the intermediate vertical sections correspond to the imaginary-time evolution that separates groups of operators in~\eqref{eq:technique-general-correlator} and~\eqref{eq:technique-correlator-interaction-picture}. The points $t = t_0$ and $t = t_0 - i (n+1) \beta$ of the contour $\mathcal{C}$ are identified, so the replicated model has a thermal initial quantum state with the effective inverse temperature $\beta_n = (n+1) \beta$.

\begin{figure}[t]
    \center{\includegraphics[width=\linewidth]{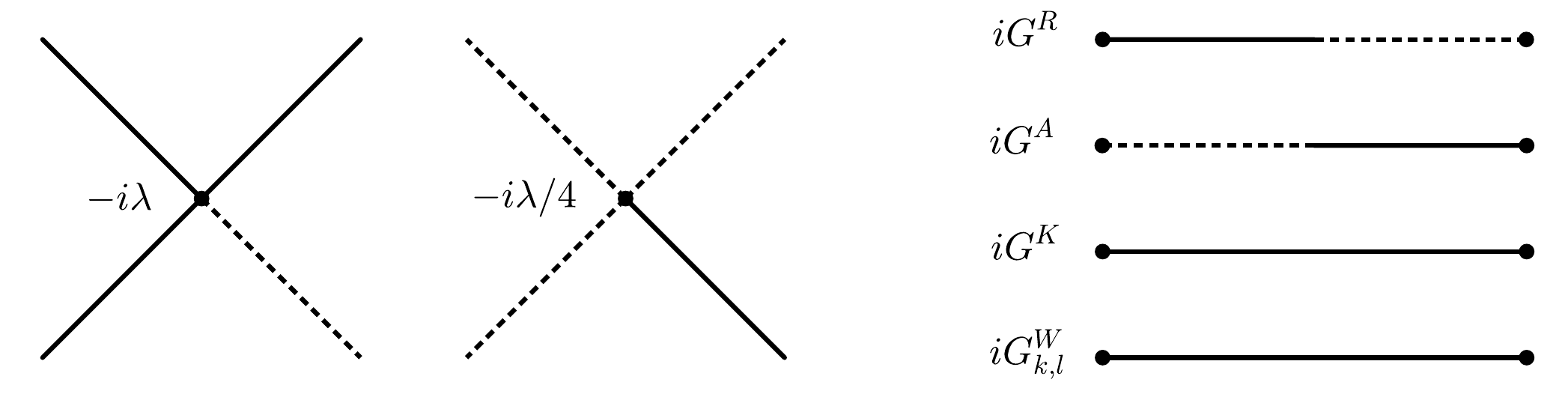}}
    \caption{Vertices~\eqref{eq:technique-cq-vertices} and propagators~\eqref{eq:technique-cq-propagators} of the Schwinger-Keldysh technique on a $2n$-fold Keldysh contour in the model~\eqref{eq:H-phi-4}. Solid and dashed lines correspond to classical and quantum components of the field, respectively. We use the same notation for $G^K$ and $G_{k,l}^W$ because the Keldysh propagator is essentially the Wightman propagator that connects points on the same fold.}
    \label{fig:diagram-rules}
\end{figure}

From the diagrammatic point of view, a perturbative expansion of the exponent in~\eqref{eq:technique-correlator-time-ordered} corresponds to a sum of all possible diagrams with $4n$ different vertices (each vertex sits on a separate horizontal branch):
\beq \label{eq:technique-pm-vertices}
\pm i \frac{\lambda}{4} \int_{t_0}^\infty \hat{x}_{k,\pm}^4(t') dt', \quad \text{where} \quad k = 1, 2, \,\cdots, 2n, \eeq
and $4 n^2$ tree-level propagators (which connect the horizontal branches):
\beq \label{eq:technique-pm-propagators}
i G_{k,\pm; l,\pm}(t,t') = \left\langle \hat{x}_{k, \pm}(t) \hat{x}_{l, \pm}(t') \right\rangle_0, \quad \text{where} \quad k, l = 1, 2, \,\cdots, 2n. \eeq
The number of all possible diagrams is enormous already at the first order of perturbative expansion and grows exponentially with powers of $\lambda$. This makes such a diagram technique unfeasible. Nevertheless, it can be simplified substantially if we notice that propagators on the same fold are linearly dependent:
\beq \label{eq:technique-propagator-dependence}
G_{k,-; k,-}(t,t') + G_{k,+; k,+}(t,t') = G_{k,-; k,+}(t,t') + G_{k,+; k,-}(t,t'). \eeq
Hence, we can make a Keldysh rotation to the so-called ``classical'' (c) and ``quantum'' (q) components on each separate fold:
\beq \label{eq:technique-rotation}
\bem \hat{x}_{k,c} \\ \hat{x}_{k,q} \eem = \bem \frac{1}{2} & \frac{1}{2} \\ 1 & -1 \eem \bem \hat{x}_{k,-} \\ \hat{x}_{k,+} \eem. \eeq
After the rotation, interaction vertices acquire the following form:
\beq \label{eq:technique-cq-vertices}
-i \lambda \int_{t_0}^\infty \hat{x}_{k,c}^3(t') \hat{x}_{k,q}(t') dt', \qquad -i \frac{\lambda}{4} \int_{t_0}^\infty \hat{x}_{k,c}(t') \hat{x}_{k,q}^3(t') dt', \eeq
and all nontrivial tree-level propagators are as follows:
\beq \label{eq:technique-cq-propagators}
\begin{aligned}
i G^R(t,t') &= \left\langle \hat{x}_{k,c}(t) \hat{x}_{k,q}(t') \right\rangle_0 = -i \theta(t - t') \, \frac{\sin \left[ m (t - t') \right]}{m}, \\
i G^A(t,t') &= \left\langle \hat{x}_{k,q}(t) \hat{x}_{k,c}(t') \right\rangle_0 = i \theta(-t + t') \, \frac{\sin \left[ m (t - t') \right]}{m}, \\
i G^K(t,t') &= \left\langle \hat{x}_{k,c}(t) \hat{x}_{k,c}(t') \right\rangle_0 = \frac{1}{2} \coth \! \left[ \frac{\beta (n + 1) m}{2} \right] \frac{\cos \left[ m (t - t') \right]}{m}, \\
i G_{k,l}^W(t,t') &= \left\langle \hat{x}_{k,c}(t) \hat{x}_{l,c}(t') \right\rangle_0 = \frac{e^{\beta \left[ n + 1 - (l - k)/2\right] m}}{e^{\beta (n + 1) m} - 1} \, \frac{e^{- i m (t - t')}}{2 m} + \frac{e^{\beta \left[ (l - k)/2\right] m}}{e^{\beta (n + 1) m} - 1} \, \frac{e^{i m (t - t')}}{2 m},
\end{aligned} \eeq
where $k \neq l$ in the Wightman (W) propagator. The rules~\eqref{eq:technique-cq-vertices} and~\eqref{eq:technique-cq-propagators} are conveniently represented in a graphical form (Fig.~\ref{fig:diagram-rules}).

We emphasize that the retarded (R), advanced (A), and Keldysh (K) propagators are the same on all folds, and --- most importantly --- the Wightman propagator, which connects different folds, contains only classical components of the fields. All other propagators are zero. In other words, quantum components of a field on any fold are \textit{not} correlated with \textit{any other} field on a different fold. This observation substantially reduces the number of allowable diagrams.

For example, let us consider the regularized ROTOC~\eqref{eq:ROTOC-reg} in the model~\eqref{eq:H-phi-4}. It is straightforward to show that in the ``cq'' notation, this correlator has the following form:
\beq \label{eq:ROTOC-phi-4}
C_n^\mathrm{reg}(t) = \left(- 
\frac{1}{2} \right)^n \left\langle 
\sum_{i_1, j_1, \cdots, i_{2n}, j_{2n}} \hat{z}^{i_1}_{1,c}(t) \hat{z}^{j_1}_{1,q}(0) \hat{z}^{i_1}_{2,c}(t) \hat{z}^{j_1}_{2,q}(0) \cdots \hat{z}^{i_{2n}}_{2n-1,c}(t) \hat{z} ^{j_{2n}}_{2n-1,q}(0) \hat{z}^{i_{2n}}_{2n,c}(t) \hat{z} ^{j_{2n}}_{2n,q}(0) \right\rangle \eeq
where $\langle \,\cdots \rangle$ is understood in the sense of~\eqref{eq:technique-correlator-time-ordered}, and upper indices run through values $i_k, j_k = 1,2$, such that $\hat{z}_{k,c/q}^1 = \hat{x}_{k,c/q}$ and $\hat{z}_{k,c/q}^2 = \hat{p}_{k,c/q}$. Correlation functions with momentum operators are derived from~\eqref{eq:technique-correlator-time-ordered} and~\eqref{eq:technique-rotation} using the relation $p^i(t) = m \dot{x}^i(t)$. At the tree level (zero order in~$\lambda$), the rules of the diagram technique, Eqs.~\eqref{eq:technique-cq-vertices} and~\eqref{eq:technique-cq-propagators}, imply that the correlator~\eqref{eq:ROTOC-phi-4} breaks down in a product of $2n$ retarded propagators:
\beq \label{eq:ROTOC-phi-4-tree}
C_n^\mathrm{reg}(t) = \left(\frac{1}{2} \sum_{i, j} G^{R;i,j}(t,0) G^{R;i,j}(t,0) \right)^n, \eeq
where
\beq i G^{R;i_k,j_k}(t,0) = \left\langle \hat{z}^{i_k}_{k,c}(t) \hat{z}^{j_k}_{k,q}(0) \right\rangle. \eeq
There are two types of loop corrections to the tree-level result~\eqref{eq:ROTOC-phi-4-tree}. First, all tree-level propagators receive loop corrections that renormalize the corresponding self-energies. These corrections are very similar to loop corrections in the standard Schwinger-Keldysh technique (which is defined on a single-fold contour) and have the same physical meaning; the only difference is that the internal vertices can belong to any fold. Second, the retarded propagators in~\eqref{eq:ROTOC-phi-4-tree} are connected by Wightman propagators. This gives a nontrivial contribution to the ROTOC. In particular, it will give an exponentially growing contribution if the kernel that generates all possible connected diagrams with $4n$ external points has a positive eigenvalue. Unfortunately, this kernel is hard to calculate in the model~\eqref{eq:H-phi-4} because there is no clear hierarchy between the contributions with a different structure or number of loops. In the following subsections, we consider several examples of models where the leading contribution to this kernel can be explicitly estimated and resummed in the limit $N \to \infty$.

\subsection{Nonlinear vector mechanics}
\label{sec:ON}

The first example of a tractable model, where the ROTOCs and the refined quantum LE can be estimated analytically, is the large-$N$ nonlinear vector mechanics --- i.e., a $(0+1)$-dimensional quantum field theory --- with an explicitly broken $O(N)$ symmetry:
\beq \label{eq:H-ON}
H = \underbrace{\sum_{i = 1}^N \left( \frac{1}{2} \dot{x}_i \dot{x}_i + \frac{m^2}{2} x_i x_i \right) + \frac{\lambda}{4 N} \sum_{i, j = 1}^N x_i x_i x_j x_j}_\text{$O(N)$-symmetric} - \underbrace{\frac{\lambda}{4 N} \sum_{\phantom{,}i = 1\phantom{,}}^N x_i x_i x_i x_i}_\text{non-symmetric}. \eeq
For convenience, we introduce the 't~Hooft coupling constant $\lambda$ and separate the $O(N)$-symmetric and non-symmetric parts of the Hamiltonian. Besides, we consider the limit $N \to \infty$ and calculate all quantites only to the leading nontrivial order in $1/N$.

Although the model~\eqref{eq:H-ON} might look artificial, it serves as a convenient toy model for many remarkable phenomena. First, this model is a large-$N$ generalization of the system of two nonlinearly coupled harmonic osillators ($H_\mathrm{int} = g x_1^2 x_2^2$), which is a paradigmatic example of both classical and quantum chaos~\cite{Pullen, Haller, Vorobev, Akutagawa}. Second, the $N = 3$ version of the Hamiltonian~\eqref{eq:H-ON} coincides with the spatially reduced $SU(2)$ Yang-Mills theory, which is a basis of the Standard Model of elementary particles~\cite{Matinyan, Chirikov, Savvidy:1984, Savvidy:2022}. Third, the model~\eqref{eq:H-ON} is very similar to a strongly coupled phonon fluid, which qualitatively describes the properties of complex oxides like SrTiO$_3$~\cite{Tulipman:2020, Tulipman:2021}. Therefore, the study of chaotic properties of the model~\eqref{eq:H-ON} might provide us with useful insights into the mentioned complex models.

\begin{figure}[t]
    \center{\includegraphics[width=\linewidth]{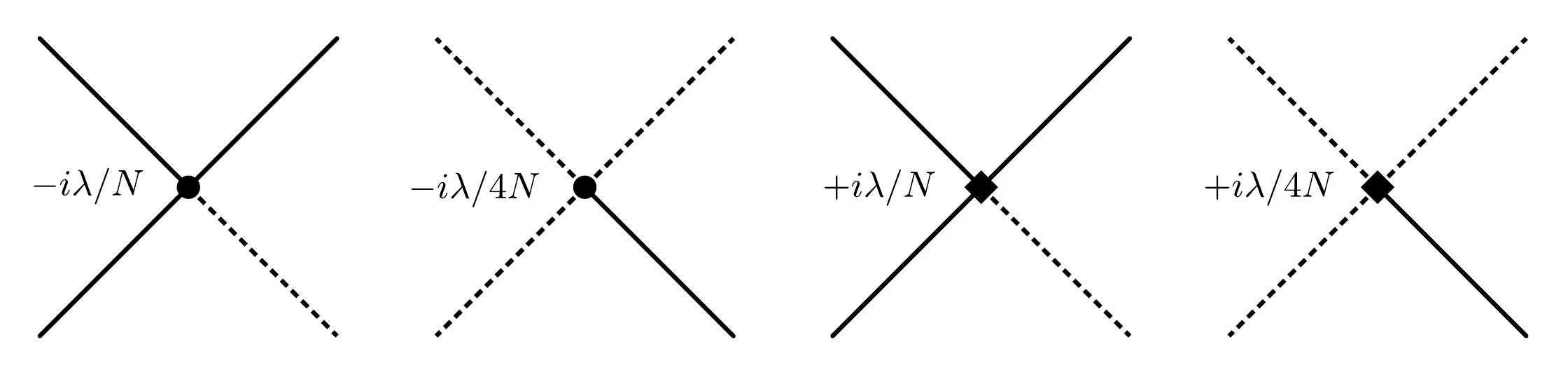}}
    \caption{$O(N)$-symmetric (circle) and non-symmetric (diamond) vertices in the model~\eqref{eq:H-ON}. The group and fold indices are suppressed for brevity.}
    \label{fig:ON-vertices}
\end{figure}

To estimate the $n$-th ROTOC in the model~\eqref{eq:H-ON}, we employ the Schwinger-Keldysh diagram technique on a $2n$-fold contour. The rules of this technique are basically the same as in the simple model~\eqref{eq:H-phi-4}. The only difference is that propagators and vertices now carry group indices:
\beq \label{eq:ON-propagators}
\begin{aligned}
\left\langle \hat{x}_{i;k,c}(t) \hat{x}_{j;k,q}(t') \right\rangle_0 &= i G^R(t,t') \delta_{i,j} = -i \theta(t - t') \, \frac{\sin \left[ m (t - t') \right]}{m} \, \delta_{i,j}, \\
\left\langle \hat{x}_{i;k,q}(t) \hat{x}_{j;k,c}(t') \right\rangle_0 &= i G^A(t,t') \delta_{i,j} = i \theta(-t + t') \, \frac{\sin \left[ m (t - t') \right]}{m} \, \delta_{i,j}, \\
\left\langle \hat{x}_{i;k,c}(t) \hat{x}_{j;k,c}(t') \right\rangle_0 &= i G^K(t,t') \delta_{i,j} = \frac{1}{2} \coth \! \left[ \frac{\beta (n + 1) m}{2} \right] \frac{\cos \left[ m (t - t') \right]}{m} \, \delta_{i,j}, \\
\left\langle \hat{x}_{i;k,c}(t) \hat{x}_{j;l,c}(t') \right\rangle_0 &= i G_{k,l}^W(t,t') \delta_{i,j} = \left[ \frac{e^{\beta \left[ n + 1 - (l - k)/2\right] m}}{e^{\beta (n + 1) m} - 1} \, \frac{e^{- i m (t - t')}}{2 m} + \frac{e^{\beta \left[ (l - k)/2\right] m}}{e^{\beta (n + 1) m} - 1} \, \frac{e^{i m (t - t')}}{2 m} \right] \delta_{i,j},
\end{aligned} \eeq
and every $O(N)$-symmetric vertex has its non-symmetric counterpart (Fig.~\ref{fig:ON-vertices}).

First of all, let us sum the loop corrections to the tree-level propagators and vertices. In the leading order in $1/N$, the non-symmetric vertices are negligible because they contain less summations over the group indices than the symmetric ones. Therefore, in this order, loop corrections to propagators and vertices are described by the standard tadpole diagrams, Fig.~\ref{fig:ON-bubbles}(a), and bubble chain diagrams, Fig.~\ref{fig:ON-bubbles}(b). Such diagrams are explicitly resummed by the corresponding Dyson-Schwinger (DS) equations. For example, the DS equation that sums the tadpole corrections to the retarded propagator has the following form:

\begin{figure}[t]
    \center{\includegraphics[width=0.75\linewidth]{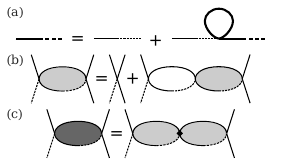}}
    \caption{(a) The DS equation that sums the leading-order loop corrections to the retarded propagator. Thin and thick lines correspond to the tree-level and resummed propagators, respectively. (b) The Dyson-Schwinger equation that sums the leading-order corrections to one of the vertices of Fig.~\ref{fig:ON-vertices}. All internal propagators are exact up to $\mO(1/N)$ corrections. The light gray loop denotes the resummed bubble chain $B(t, t')$. (c) The leading non-symmetric correction to the same vertex.}
    \label{fig:ON-bubbles}
\end{figure}

\beq \label{eq:ON-DS-R}
\tilde{G}^R(t, t') = G^R(t, t') + i \lambda \int_{t_0}^\infty dt'' \, G^R(t, t'') \tilde{G}^K(t'', t'') \tilde{G}^R(t'', t'), \eeq
where $\tilde{G}$ denotes the resummed propagators. The DS equations that resum tadpole corrections to other propagators have the same form. The solutions to these equations --- the $\mO(1)$ resummed propagators --- coincide with the tree-level propagators~\eqref{eq:ON-propagators} with a shifted mass determined by the following trascendental equation (we remind that $\beta_n = (n+1) \beta$):
\beq \label{eq:ON-m}
\frac{\tilde{m}_n^2}{m^2} = 1 + \frac{\lambda}{2 m^3} \frac{m}{\tilde{m}_n} \coth\left( \frac{\beta_n m}{2} \frac{\tilde{m}_n}{m} \right). \eeq
The DS equation that sums the bubble chain diagrams:
\beq \label{eq:ON-DS-bubble}
B(t, t') = \delta(t - t') + 2 i \lambda \int_{t_0}^\infty dt'' \, \tilde{G}^R(t, t'') \tilde{G}^K(t, t'') B(t'', t'), \eeq
is also explicitly solved using the following ansatz inspired by the structure of a single bubble: 
\beq \label{eq:ON-B}
B(t, t') = \delta(t - t') - \nu_n \, \tilde{m}_n \, \theta(t - t') \sin\left[ \mu_n \tilde{m}_n (t - t') \right], \eeq
where the coefficients $\mu_n$ and $\nu_n$ are expressed in terms of the $\mO(1)$ resummed mass:
\beq \label{eq:ON-mu-nu}
\mu_n^2 = 6 - 2 m^2 / \tilde{m}_n^2, \qquad \nu_n = \mu_n - 4 / \mu_n. \eeq
Note that in the high-temperature and weak-coupling limit:
\beq \label{eq:ON-limit}
\beta m \ll \lambda / m^3 \ll 1, \eeq
the expressions for the resummed mass~\eqref{eq:ON-m} and the coefficients~\eqref{eq:ON-mu-nu} are significantly simplified\footnote{If the replica number is large and $\beta (n + 1) m \gg 1$, the expression for the resummed mass is even simpler: $\tilde{m}_n/m \approx 1 + \lambda / 4 m^3$. However, we are mainly interested in the limit $n \to 0$ to use the replica trick~\eqref{eq:replica-trick}.}:
\beq \label{eq:ON-m-mu-nu-simple}
\tilde{m}_n \approx \sqrt[4]{\lambda / \beta_n}, \qquad \mu_n \approx \sqrt{6}, \qquad \nu_n \approx \sqrt{2/3}. \eeq
Hence, it is convenient to assume the limit~\eqref{eq:ON-limit} to get simple illustrative estimates.

Then we make the following important observation. When we try to resum the leading in $1/N$ loop corrections to the ROTOC in the model~\eqref{eq:H-ON}, we neglect the non-symmetric vertices and in fact, effectively end up with the fully $O(N)$-symmetric model. At the classical level, this model has a $2N$-dimensional phase space and exactly $N$ conserved quantities: energy $H$ and $N-1$ Casimir operators, $L_k^2 = \sum_{j=2}^{k+1} \sum_{i=1}^{j-1} L_{ij}^2$, where $k=1, \,\cdots, N-1$, $L_{ij} = x_i p_j - p_i x_j$ are the angular momenta, and $p_i = \dot{x}_i$ are the canonical momenta. Hence, the $O(N)$-symmetric model is integrable and has zero classical LE, $\bar{\kappa}_{cl} = 0$. Keeping in mind the correspondence between the average classical LE and refined quantum LE, we obtain that $\bar{\kappa}_q = \bar{\kappa}_{cl} = 0$. Hence, the LOTOC in the $O(N)$-symmetric version of~\eqref{eq:H-ON} grows slower than linearly, and the ROTOCs grow slowly than exponentially. Of course, this behavior can be established directly from the Schwinger-Keldysh technique on a $2n$-fold Keldysh contour; e.g., see the proof for the $n = 1$ case~\cite{Kolganov}. Thus, if we want to capture the chaotic behavior of the model~\eqref{eq:H-ON}, we need to include non-symemtric vertices into the kernels that generate the resummed ROTOCs.

\begin{figure}[t]
    \center{\includegraphics[width=0.8\linewidth]{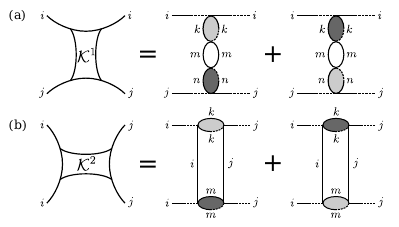}}
    \caption{The building blocks of the kernels that generate the ROTOCs. Note that we explicitly show group indices, but suppress fold indices.}
    \label{fig:ON-kernels}
\end{figure}

Since each non-symmetric vertex assigns an extra $1/N$ factor to a diagram, the leading non-symmetric correction to a generating kernel is obtained by replacing as less as possible $O(N)$-symmetric vertices with their non-symmetric analogs. This procedure modifies the corresponding bubble chains and is described by the diagram Fig.~\ref{fig:ON-bubbles}(c):
\beq \label{eq:ON-bubble-nonsymm}
\begin{aligned}
B(t, t') \to \tilde{B}(t, t') = -\frac{3}{N} \delta(t - t') &+ \frac{3}{N} \frac{\left( \mu_n^2 - 4 \right) \left( 3 \mu_n^2 + 4 \right)}{2 \mu_n^3} \, \theta\!\left( t - t' \right) \tilde{m}_n \sin \left[ \mu_n \tilde{m}_n (t - t') \right] \\ &+ \frac{3}{N} \frac{\left( \mu_n^2 - 4 \right)^2}{2 \mu_n^2} \, \theta\!\left( t - t' \right) (t - t') \cos \left[ \mu_n \tilde{m}_n (t - t') \right].   
\end{aligned} \eeq
As a result, we obtain that the kernels generating the ROTOCs are build from the following four-point blocks, see Fig.~\ref{fig:ON-kernels}:
\beq \label{eq:ON-kernels}
\begin{aligned}
\mK^1_{k,l}(t_1, t_1'; t_2, t_2') &= -\frac{8 \lambda^2}{N} \int dt_1'' dt_2'' \, \tilde{G}^R(t_1, t_1') \tilde{G}^R(t_2, t_2') B(t_1', t_1'') \tilde{B}(t_2', t_2'') \tilde{G}^W_{k,l}(t_1'', t_2'') \tilde{G}^W_{k,l}(t_1'', t_2'') \\ &\phantom{=} -\frac{8 \lambda^2}{N} \int dt_1'' dt_2'' \, \tilde{G}^R(t_1, t_1') \tilde{G}^R(t_2, t_2') \tilde{B}(t_1', t_1'') B(t_2', t_2'') \tilde{G}^W_{k,l}(t_1'', t_2'') \tilde{G}^W_{k,l}(t_1'', t_2''), \\
\mK^2_{k,l}(t_1, t_1'; t_2, t_2') &= -\frac{4 \lambda^2}{N} \int dt_1'' dt_2'' \, \tilde{G}^R(t_1, t_1'') \tilde{G}^R(t_2, t_2'') \tilde{G}^W_{k,l}(t_1'', t_2'') B(t_1'', t_1') \tilde{B}(t_2'', t_2') \tilde{G}^W_{k,l}(t_1', t_2') \\ &\phantom{=} -\frac{4 \lambda^2}{N} \int dt_1'' dt_2'' \, \tilde{G}^R(t_1, t_1'') \tilde{G}^R(t_2, t_2'') \tilde{G}^W_{k,l}(t_1'', t_2'') \tilde{B}(t_1'', t_1') B(t_2'', t_2') \tilde{G}^W_{k,l}(t_1', t_2').
\end{aligned} \eeq

We emphasize that both building blocks~\eqref{eq:ON-kernels} have an eigenvalue with a positive real part. On the contrary, fully symmetric analogs of these building blocks (which have the same structure as Eqs.~\eqref{eq:ON-kernels}, but do not contain bubble chains~$\tilde{B}$) have purely imaginary eigenvalues, see the $n = 1$ calculation~\cite{Kolganov}. Hence, if we include one or more fully symmetric blocks into the generating kernel of the ROTOC, we will get subleading corrections to the ROTOCs and replica LEs. 

\begin{figure}[t]
    \center{\includegraphics[width=0.75\linewidth]{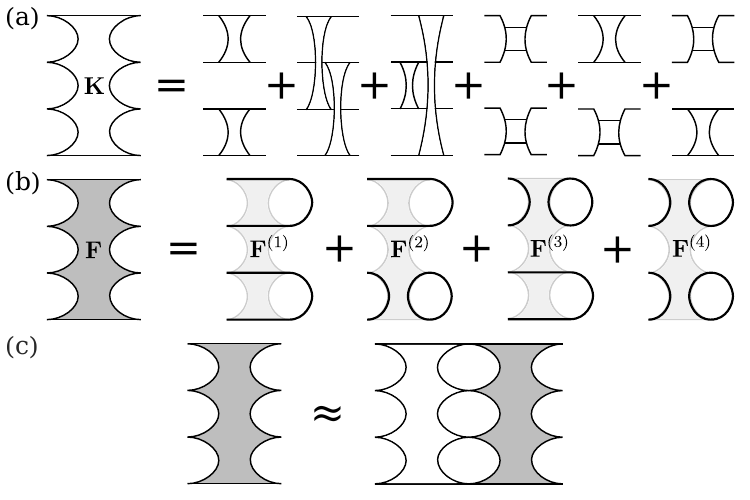}}
    \caption{(a) The kernel that generates the leading nontrivial corrections to the $C_2^\mathrm{reg}$, i.e., captures the leading in $1/N$ contribution to the second replica LE $\kappa_2$. (b) Group structure of the resummed correlator $\mathbf{F}$. Bold lines mark the chains of group indices contractions. (c) The approximate DS equation that sums ladder diagrams, i.e, the leading nontrivial loop corrections to the ROTOC.}
    \label{fig:ON-C2}
\end{figure}

In particular, let us build the generating kernel and calculate the resummed ROTOC for a simple example of $C_2^\mathrm{reg}(t)$. For convenience, we rewrite the ROTOC:
\beq \label{eq:ON-C2}
C_2^\mathrm{reg}(t) = \lim_{\substack{t_1, t_2, t_3, t_4 \; \to \; t \\ t_1', t_2', t_3', t_4' \; \to \; 0}} \left[ \prod_{\alpha = 1}^8 \frac{1 + m \pd_{t_\alpha}}{\sqrt{2}} \right] \sum_{i,i'} F_{i,i,i',i'}(t_1, t_1'; t_2, t_2'; t_3, t_3'; t_4, t_4'), \eeq
using an auxiliary correlation function $\mathbf{F} = F_{i_1,i_2,i_3,i_4}(t_1, t_1'; t_2, t_2'; t_3, t_3'; t_4, t_4')$, which contains only coordinate operators:
\beq \label{eq:ON-F}
\mathbf{F} = \sum_{j, j'} \Big\langle x_{i_1; 1, c}(t_1) \, x_{j; 1, q}(t_1') \, x_{i_2; 2, c}(t_2) \, x_{j; 2, q}(t_2') \, x_{i_3; 3, c}(t_3) \, x_{j'; 3, q}(t_3') \, x_{i_4; 4, c}(t_4) \, x_{j'; 4, q}(t_4') \Big\rangle. \eeq
Similarly to the simple model~\eqref{eq:H-phi-4}, at the tree level, the correlation function~\eqref{eq:ON-F} breaks down in a product of retarded propagators sitting on different folds of the Keldysh contour:
\beq \label{eq:ON-F_0}
\mathbf{F}_0 = \frac{1}{N^2} \delta_{i_1, i_2} \delta_{i_3, i_4} \tilde{G}^R(t_1, t_1') \tilde{G}^R(t_2, t_2') \tilde{G}^R(t_3, t_3') \tilde{G}^R(t_4, t_4'). \eeq
The lowest-order-in-$1/N$ corrections to this tree-level correlator are given by ladder-type diagrams, which are generated by the following kernel:
\beq \label{eq:ON-K}
\mathbf{K} = \mathbf{K}^{(1)} + \mathbf{K}^{(2)} + \mathbf{K}^{(3)} + \mathbf{K}^{(4)} + \mathbf{K}^{(5)} + \mathbf{K}^{(6)}, \eeq 
where we explicitly separated contributions with a different structure of group indices, see Fig.~\ref{fig:ON-C2}(a):
\beq \label{eq:ON-K-parts} \begin{aligned}
\mathbf{K}^{(1)} &= \delta_{i_1,j_1} \delta_{i_2,j_2} \delta_{i_3,j_3} \delta_{i_4,j_4} \mK_{1,2}^1(t_1,t_1';t_2,t_2') \mK_{3,4}^1(t_3,t_3';t_4,t_4'), \\
\mathbf{K}^{(2)} &= \delta_{i_1,j_1} \delta_{i_2,j_2} \delta_{i_3,j_3} \delta_{i_4,j_4} \mK_{1,3}^1(t_1,t_1';t_3,t_3') \mK_{2,4}^1(t_2,t_2';t_4,t_4'), \\
\mathbf{K}^{(3)} &= \delta_{i_1,j_1} \delta_{i_2,j_2} \delta_{i_3,j_3} \delta_{i_4,j_4} \mK_{1,4}^1(t_1,t_1';t_4,t_4') \mK_{2,3}^1(t_2,t_2';t_3,t_3'), \\
\mathbf{K}^{(4)} &= \delta_{i_1,i_2} \delta_{i_3,i_4} \delta_{j_1,j_2} \delta_{j_3,j_4} \mK_{1,2}^2(t_1,t_1';t_2,t_2') \mK_{3,4}^2(t_3,t_3';t_4,t_4'), \\
\mathbf{K}^{(5)} &= \delta_{i_1,j_1} \delta_{i_2,j_2} \delta_{i_3,i_4}  \delta_{j_3,j_4} \mK_{1,2}^1(t_1,t_1';t_2,t_2') \mK_{3,4}^2(t_3,t_3';t_4,t_4'), \\
\mathbf{K}^{(6)} &= \delta_{i_1,i_2} \delta_{j_1,j_2} \delta_{i_3,j_3} \delta_{i_4,j_4} \mK_{1,2}^2(t_1,t_1';t_2,t_2') \mK_{3,4}^1(t_3,t_3';t_4,t_4').
\end{aligned} \eeq
Essentially, we use that the building block $\mK^1_{k,l}$ simply extends the lines it is attached to, whereas $\mK^2_{k,l}$ closes a loop and opens another at the opposite side (and gives a negligible, suppressed by powers of $1/N$, contribution if it does not close a loop).

Applying the kernel~\eqref{eq:ON-K} to the tree-level correlation function~\eqref{eq:ON-F_0}, we see that the resummed version of this correlator inherits a particular structure of group indices depicted at Fig.~\ref{fig:ON-C2}(b):
\beq \label{eq:ON-F-structure}
\mathbf{F} = \delta_{i_1, i_2} \delta_{i_3, i_4} F(t_1,t_1';t_2,t_2';t_3,t_3';t_4,t_4') = \mathbf{F}^{(1)} + \mathbf{F}^{(2)} + \mathbf{F}^{(3)} + \mathbf{F}^{(4)}. \eeq
In fact, different parts of the kernel~\eqref{eq:ON-K-parts} move the correlators~\eqref{eq:ON-F-structure} between classes with a different structure of group indices. So, the DS equation that sums the leading nontrivial loop corrections to these correlators have the following form: 
\beq \label{eq:ON-DS-parts} \begin{gathered}
\mathbf{F}^{(1)} = \mathbf{F}_0 + \left[ \mathbf{K}^{(1)} + \mathbf{K}^{(2)} + \mathbf{K}^{(3)} \right] \circ \mathbf{F}^{(1)}, \\
\mathbf{F}^{(2)} = \mathbf{K}^{(5)} \circ \mathbf{F}^{(1)} + \left[ \mathbf{K}^{(1)} + \mathbf{K}^{(2)} + \mathbf{K}^{(3)} + \mathbf{K}^{(5)} \right] \circ \mathbf{F}^{(2)}, \\
\mathbf{F}^{(3)} = \mathbf{K}^{(6)} \circ \mathbf{F}^{(1)} + \left[ \mathbf{K}^{(1)} + \mathbf{K}^{(2)} + \mathbf{K}^{(3)} + \mathbf{K}^{(6)} \right] \circ \mathbf{F}^{(3)}, \\
\mathbf{F}^{(4)} = \mathbf{K}^{(4)} \circ \mathbf{F}^{(1)} + \left[ \mathbf{K}^{(4)} + \mathbf{K}^{(6)} \right] \circ \mathbf{F}^{(2)} + \left[ \mathbf{K}^{(4)} + \mathbf{K}^{(5)} \right] \circ \mathbf{F}^{(3)} + \mathbf{K} \circ \mathbf{F}^{(4)},
\end{gathered} \eeq
where we introduce a short notation for the operator that adds one extra kernel from Fig.~\ref{fig:ON-C2}(a) to a diagram:
\beq \label{eq:ON-add-a-rung}
\mathbf{K} \circ \mathbf{F} \equiv \int d t_1' d t_2' d t_3' d t_4' \sum_{j_1,j_2,j_3,j_4} K_{i_1,i_2,i_3,i_4}^{j_1,j_2,j_3,j_4}(t_1,t_1';t_2,t_2';t_3,t_3';t_4,t_4') F_{j_1,j_2,j_3,j_4}(t_1',t_1'';t_2',t_2'';t_3',t_3'';t_4',t_4''). \eeq
Summing Eqs.~\eqref{eq:ON-DS-parts}, we get the DS equation on the full correlator, see Fig.~\ref{fig:ON-C2}(c):
\beq \label{eq:ON-DS-F}
\mathbf{F} = \mathbf{F}_0 + \mathbf{K} \circ \mathbf{F} \approx \mathbf{K} \circ \mathbf{F}. \eeq
For the last identity, we assumed that the kernel $\mathbf{K}$ has an eigenvalue with a positive real part, so the solution to Eq.~\eqref{eq:ON-DS-F} grows exponentially, and the bounded contribution of $\mathbf{F}_0$ is negligible at large times. More precisely, we propose the following ansatz:
\beq \label{eq:ON-ansatz}
F(t_1,t_1';t_2,t_2';t_3,t_3';t_4,t_4') \sim \exp\left[ \varkappa_2 \left( t_1 + t_2 + t_3 + t_4 - t_1' - t_2' - t_3' - t_4' \right) \right], \eeq
which is inspired by the expected exponential growth of $\mathbf{F}$ and the fact that all times enter this correlator on equal terms, see the definition~\eqref{eq:ON-F}. Substituting the ansatz into the DS equation~\eqref{eq:ON-DS-F}, we obtain the equation on $\varkappa_2$:
\small \beq \label{eq:ON-replica-LE-eq} \begin{aligned}
1 &\approx 3 \times\! \left[ - \frac{1536}{N^2} \frac{\lambda^2 w_2^2}{\left( \mu_2 \tilde{m}_2 \right)^6} \frac{1}{\left( 1 + \frac{\varkappa_2^2}{\tilde{m}_2^2} \right)^2} \right]^2 \!\!+ \left[\!\vphantom{\frac{1536}{N^2} \frac{\lambda^2 w_2^2}{\left( \mu_2 \tilde{m}_2 \right)^6} \frac{1}{\left( 1 + \frac{\varkappa_2^2}{\tilde{m}_2^2} \right)^2}} - \frac{24}{N^2} \frac{\lambda^2 w_2^2}{\tilde{m}_2^6} \frac{A_2\left( \varkappa_2 / \tilde{m}_2 \right)}{\left( 1 + \frac{\varkappa_2^2}{\tilde{m}_2^2} \right) \left( (\mu_2 - 1)^2 + \frac{\varkappa_2^2}{\tilde{m}_2^2} \right)^2 \left( (\mu_2 + 1)^2 + \frac{\varkappa_2^2}{\tilde{m}_2^2} \right)^2 } \right]^2 \\ &\,+ 2 \times\! \left[ - \frac{1536}{N^2} \frac{\lambda^2 w_2^2}{\left( \mu_2 \tilde{m}_2 \right)^6} \frac{1}{\left( 1 + \frac{\varkappa_2^2}{\tilde{m}_2^2} \right)^2} \right] \times \left[ \!\vphantom{\frac{1536}{N^2} \frac{\lambda^2 w_2^2}{\left( \mu_2 \tilde{m}_2 \right)^6} \frac{1}{\left( 1 + \frac{\varkappa_2^2}{\tilde{m}_2^2} \right)^2}} - \frac{24}{N^2} \frac{\lambda^2 w_2^2}{\tilde{m}_2^6} \frac{A_2\left( \varkappa_2 / \tilde{m}_2 \right)}{\left( 1 + \frac{\varkappa_2^2}{\tilde{m}_2^2} \right) \left( (\mu_2 - 1)^2 + \frac{\varkappa_2^2}{\tilde{m}_2^2} \right)^2 \left( (\mu_2 + 1)^2 + \frac{\varkappa_2^2}{\tilde{m}_2^2} \right)^2 } \right],
\end{aligned} \eeq \normalsize
where we introduce a short notation for the coefficient
\beq w_2 \equiv \frac{e^{\beta_2 \tilde{m}_2 / 2}}{e^{\beta_2 \tilde{m}_2} - 1} \eeq
and the function
\beq \begin{aligned}
A_2(\varkappa) &= 2 \left( \varkappa^2 + 9 \right) \left[ \varkappa^4 + \varkappa^2 \left( \mu_2^2 + 6 \right) + 3 \left( \mu_2^2 - 1 \right) \right] \\ &\phantom{=}+ e^{- 2 \beta \tilde{m}_2} \left( \varkappa^2 + 1 \right) \left( \varkappa - i \right) \left( \varkappa + 3 i \right) \left[ \left( \varkappa - i \right)^2 + \mu_2^2 \right] \\
&\phantom{=}+ e^{+ 2 \beta \tilde{m}_2} \left( \varkappa^2 + 1 \right) \left( \varkappa + i \right) \left( \varkappa - 3 i \right) \left[ \left( \varkappa + i \right)^2 + \mu_2^2  \right].
\end{aligned} \eeq
The ``$\approx$'' in Eq.~\eqref{eq:ON-replica-LE-eq} emphasizes that we neglect the higher-order corrections in $1/N$ (including the corrections that comprise fully symmetric building blocks). The first term on the r.h.s. of Eq.~\eqref{eq:ON-replica-LE-eq} emerges from the kernels $\mathbf{K}^{(1)}$, $\mathbf{K}^{(2)}$, and $\mathbf{K}^{(3)}$, which are built purely from $\mK^1_{k,l}$. Explicitly taking the time integrals, we find that all these kernels give equal contributions to Eq.~\eqref{eq:ON-replica-LE-eq}; furthermore, surprisingly, these contributions do not depend on the fold indices $k$ and $l$. This fact is summarized by a simple combinatorial factor of $3$ in front of the first term. The second term emerges from the kernel $\mathbf{K}^{(4)}$, and the last term emerges from the kernels $\mathbf{K}^{(5)}$ and $\mathbf{K}^{(6)}$. The dependence of these terms on the fold indices result in the factors $e^{\pm 2 \beta \tilde{m}_2}$ in the function $A_2(t)$.

To solve Eq.~\eqref{eq:ON-replica-LE-eq} with respect to $\varkappa_2$, we notice that expressions in the square brackets are multiplied by small factors of $1/N^2$ and have double poles at the points $\varkappa_2 = \pm i \tilde{m}_2$, $\varkappa_2 = \pm i (\mu_2 - 1) \tilde{m}_2$, and $\varkappa_2 = \pm i (\mu_2 + 1) \tilde{m}_2$. Hence, the solution to the equation has a form $\varkappa_2 = P + R/N + \mO\left( 1/N^2 \right)$, where $P$ is a pole and $R$ is a coefficient that depends on $\lambda$, $\tilde{m}_2$, $\mu_2$ and $w_2$, but does not depend on $\varkappa_2$ and $N$. Substituting such approximate solutions to Eq.~\eqref{eq:ON-replica-LE-eq} and keeping only the leading-order-in-$1/N$ terms, we estimate the coefficients $R$ near different poles. Finally, we select the solutions with the largest positive real part, which are as follows:
\beq \label{eq:ON-replica-LE-sol}
\varkappa_2 \approx \pm i \tilde{m}_2 + 3^{1/4} \times \frac{8 \sqrt{6}}{N} \frac{\lambda w_2}{(\mu_2 \tilde{m}_2)^3} \tilde{m}_2. \eeq
In fact, this solution emerges purely from the first term of Eq.~\eqref{eq:ON-replica-LE-eq} because other terms give lower-order corrections in $1/N$ near the poles $\kappa_2 = \pm i \tilde{m}_2$. In other words, if we need to capture only the leading nonzero contributions to the positive real part of $\kappa_2$, we can practically keep only the kernels built from $\mK_{k,l}^1$ (first three kernels at Fig.~\ref{fig:ON-C2}(a)) and solve the following truncated equation instead of Eq.~\eqref{eq:ON-replica-LE-eq}:
\beq \label{eq:ON-replica-LE-eq-truncated}
1 \approx 3 \times\! \left[ - \frac{1536}{N^2} \frac{\lambda^2 w_2^2}{\left( \mu_2 \tilde{m}_2 \right)^6} \frac{1}{\left( 1 + \varkappa_2^2 / \tilde{m}_2^2 \right)^2} \right]^2. \eeq

To extend Eqs.~\eqref{eq:ON-replica-LE-eq} and~\eqref{eq:ON-replica-LE-eq-truncated} to other replica numbers, we make two observations. First, we trace the origin of expressions in square brackets in Eq.~\eqref{eq:ON-replica-LE-eq} to the building blocks~\eqref{eq:ON-kernels}. In other words, we notice that the generating kernel of the $n$-th ROTOC --- a generalization of Eqs.~\eqref{eq:ON-K} and~\eqref{eq:ON-K-parts}, --- is built from building blocks $\mK^1_{k,l}$ and $\mK^2_{k,l}$ similarly to its $n = 2$ version. Then we substitute an exponentially growing ansatz to the DS equation that sums the leading nontrivial loop corrections to the ROTOC similarly to~\eqref{eq:ON-DS-F}:
\beq \label{eq:ON-ansatz-n}
\mathbf{F} \sim \exp\left[ \varkappa_n \sum_{a = 1}^{2n} \left( t_a - t_a' \right) \right]. \eeq
As a result, each building block $\mK^1_{k,l}$ that constitutes a generating kernel leads to an expression with double poles $\varkappa_n = \pm i \tilde{m}_n$, and each $\mK^2_{k,l}$ leads to an expression with double poles $\varkappa_n = \pm i (\mu_n - 1) \tilde{m}_n$ and $\varkappa_n = \pm i (\mu_n + 1) \tilde{m}_n$. Thus we obtain an analog of Eq.~\eqref{eq:ON-replica-LE-eq}.

Second, similarly to the case $n = 2$, the analog of Eq.~\eqref{eq:ON-replica-LE-eq} has solutions near all the mentioned poles. Furthermore, the solutions near the poles $\varkappa_n = \pm i \tilde{m}_n$ have a larger real positive part than all other solutions, at least in the limit~\eqref{eq:ON-limit}\footnote{On the one hand, for a small $n$, the dominance of solutions $\varkappa_n \approx \pm i \tilde{m}_n$ can be checked directly. On the other hand, for a large $n$, we can prove that all other solutions are parametrically suppressed in the limit~\eqref{eq:ON-limit}.}. Therefore, we can keep only the first term in the equation and arrive at the analog of Eq.~\eqref{eq:ON-replica-LE-eq-truncated}:
\beq \label{eq:ON-replica-LE-n-eq}
1 \approx (2n - 1)!! \times\! \left[ - \frac{1536}{N^2} \frac{\lambda^2 w_n^2}{\left( \mu_n \tilde{m}_n \right)^6} \frac{1}{\left( 1 + \varkappa_n^2 / \tilde{m}_n^2 \right)^2} \right]^n, \eeq
where $w_n \equiv e^{\beta_n \tilde{m}_n / 2} \big/ \left( e^{\beta_n \tilde{m}_n} - 1 \right)$.

The combinatorial factor of $(2n-1)!!$ in Eq.~\eqref{eq:ON-replica-LE-n-eq} counts all possible ways to connect the retarded propagators on a $2n$-fold contour by the building blocks $\mK^1_{k,l}$, all of which lead to the contributions of the form~\eqref{eq:ON-replica-LE-n-eq}.

Finally, solving Eq.~\eqref{eq:ON-replica-LE-n-eq}, substituting the solution to the ansatz~\eqref{eq:ON-ansatz-n}, and taking in mind the generalization of~\eqref{eq:ON-C2} to arbitrary $n$, we estimate the ROTOC:
\beq C_n^\mathrm{reg}(t) \sim e^{2 \kappa_n t}, \eeq
where the replica LE $\kappa_n$ is as follows:
\beq \label{eq:ON-replica-LE}
\kappa_n = n \varkappa_n \approx n \left[(2n-1)!!\right]^{\frac{1}{2n}} \frac{8 \sqrt{6}}{N} \frac{\lambda \tilde{m}_n}{(\mu_n \tilde{m}_n)^3} \frac{e^{\beta_n \tilde{m}_n / 2}}{e^{\beta_n \tilde{m}_n} - 1}. \eeq
Analytically continuing this expression to real $n$ and using the replica trick~\eqref{eq:replica-trick}, we obtain the LOTOC:
\beq \label{eq:ON-LOTOC}
L(t) \approx 2 \bar{\kappa}_q t + \const, \eeq
and the refined quantum LE:
\beq \label{eq:ON-true-qLE}
\bar{\kappa}_q \approx \sqrt{2} e^{\psi(1/2)/2} \times \frac{8 \sqrt{6}}{N} \frac{\lambda \tilde{m}}{(\mu \tilde{m})^3} \frac{e^{\beta \tilde{m} / 2}}{e^{\beta \tilde{m}} - 1}, \eeq
where $\psi(x) = \Gamma'(x)/\Gamma(x)$ is the digamma function, and $\tilde{m}$ and $\mu$ are determined by Eqs.~\eqref{eq:ON-m} and~\eqref{eq:ON-mu-nu} with $n = 0$ and $\beta_0 = \beta$. In the limit~\eqref{eq:ON-limit}, this expression is additionally simplified:
\beq \label{eq:ON-true-qLE-simplified-high}
\bar{\kappa}_q \approx 0.7 \, \frac{1}{N} \, \sqrt[4]{\frac{\lambda}{\beta}}. \eeq
Besides, the expression for the refined quantum LE~\eqref{eq:ON-true-qLE} is simplified in the low-temperature limit, $\beta m \gg 1$, where correlations between different folds (i.e., the Wightman propagators) are exponentially suppressed:
\beq \label{eq:ON-true-qLE-simplified-low}
\bar{\kappa}_q \approx 1.3 \, \frac{1}{N} \, \frac{\lambda}{m^3} \, e^{-\beta m / 2} \, m, \eeq
where we assume $\lambda / m^3 \ll 1$ for simplicity. This suppression probably reflects the presence of the Kolmogorov-Arnold-Moser island in the classical version of the model~\eqref{eq:H-ON}. Indeed, at low temperatures, the semiclassical averaging in~\eqref{eq:LOTOC-def} is effectively performed only over a small region near the global minimum of the total energy (roughly speaking, only the ground state contributes to the averaging). The energy of the model~\eqref{eq:H-ON} has the global minimum at the origin, $\mathbf{x} = 0$ and $\mathbf{p} = 0$, and the model is approximately free near this point. Hence, only the exponentially short boundary of this region contributes to the semiclassical LOTOC, so the average classical LE and the refined quantum LE are exponentially suppressed at small temperatures (see also the numerical calculations~\cite{Kolganov, companion}).

We also emphasize that the refined quantum LE~\eqref{eq:ON-true-qLE} is approximately two times smaller than the conventional quantum LE, which is extracted from the ordinary OTOC~\eqref{eq:OTOC-def}. From the diagrammatic point of view, this difference reflects a complex structure of correlations between different folds of the Keldysh contour. In particular, it is easy to see that the resummed ROTOC does not factorize into $n=1$ expressions even at the $n=2$ level, see Fig.~\ref{fig:ON-C2}.

\subsection{Sachdev-Ye-Kitaev model}
\label{sec:SYK}

Another important example of a chaotic many-body system is the Sachdev-Ye-Kitaev (SYK) model, which describes a $(0+1)$-dimensional system of $N$ Majorana fermions with all-to-all random interactions~\cite{Polchinski, Maldacena-SYK, Kitaev, Sarosi, Rosenhaus, Trunin-SYK}:
\beq \label{eq:H-SYK}
H = \frac{i^{q/2}}{q!} \sum_{i_1, i_2, \cdot\cdot\cdot, i_q = 1}^N j_{i_1, i_2, \cdot\cdot\cdot, i_q} \, \chi_{i_1} \chi_{i_2} \cdots \chi_{i_1}, \eeq
where operators $\chi_i$ have the following commutation relations:
\beq \label{eq:SYK-chi}
\left\{ \chi_i, \chi_j \right\} = \delta_{i,j} \quad \text{for all} \quad i,j = 1, 2, \cdot\cdot\cdot, N, \eeq
and the coupling constants are randomly drawn from a Gaussian distribution with zero mean and the following variance:
\beq \label{eq:SYK-J}
\overline{j_{i_1, i_2, \cdot\cdot\cdot, i_q}^2} = \frac{J^2 (q-1)!}{N^{q-1}} = \frac{2^{q-1}}{q} \frac{\mathcal{J}^2 (q-1)!}{N^{q-1}} \qquad \text{(no sum assumed)}. \eeq
Note that the Gaussianity of the couplings~\eqref{eq:SYK-J} implies that the disorder average of an odd number of couplings is zero, and the average of an even number of couplings split into a sum of all possible two-coupling disorder averages. In what follows, we always consider the disorder-averaged quantities (propagators, ROTOCs, etc.).

We emphasize that the model~\eqref{eq:H-SYK} has many remarkable properties and qualitatively describes a variety of important phenomena. First, it is the simplest model of quantum holography: in the infrared limit, the $(0+1)$-dimensional SYK model is dual to the $(1+1)$-dimensional Jackiw-Teitelboim gravity in the AdS$_2$ space. Hence, the SYK model can be used as a toy model of many phenomena related to black holes and other curved spaces, such as information and cloning paradoxes~\cite{Hayden, Sekino}, black hole --- wormhole phase transition~\cite{Qi, Garcia-Garcia, Milekhin:2019}, or teleportation through a traversable wormhole~\cite{Yoshida, Gao:2019, Milekhin:2022, Jafferis:Nature}. Second, the SYK model qulitatively describes the physics of strange metals and non-fermi liquids~\cite{Hartnoll, Sachdev:2015, Song, Chowdhury:2021}. Third, this model is both solvable in the large-$N$ limit, convenient for numerical calculations, and maximally chaotic, which makes it a perfect model to study quantum chaos, information scrambling, and related effects~\cite{Rosenhaus, Sarosi, Trunin-SYK}. Finally, the SYK model can be experimentally implemented using graphene flakes~\cite{Franz, Kruchkov, Brzezinska}, cold atoms~\cite{Danshita, Wei}, cavity quantum electrodynamics~\cite{Uhrich}, quantum dots~\cite{Chew}, or quantum computers~\cite{Jafferis:Nature, Luo, Garcia-Alvarez, Babbush}. This provides us with a unique opportunity to test our neat theoretical concepts and calculations in a real-world setup, which further underlines the importance of the SYK model.

\begin{figure}[t]
    \center{\includegraphics[width=\linewidth]{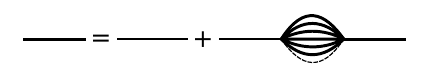}}
    \caption{The DS equation that sums the melonic (leading in $1/N$) loop corrections to the tree-level Euclidean propagator in the SYK model with $q = 7$. Thin and thick lines correspond to the tree-level and resummed propagators, respectively. Dashed line denotes the contraction of the corresponding coupling constants using~\eqref{eq:SYK-J}.}
    \label{fig:SYK-propagator}
\end{figure}

To estimate the ROTOC $C_n^\mathrm{reg}(t)$ and the refined quantum LE in the model~\eqref{eq:H-SYK}, we first consider its Euclidean version and study loop corrections to the propagators. It is easy to show that in the limit $N \to \infty$, the diagram expansion of the exact Euclidean propagator is given by the so-called melonic diagrams (Fig.~\ref{fig:SYK-propagator}), so the DS equation on the resummed propagator $G(\tau)$ has the following form~\cite{Polchinski, Maldacena-SYK, Kitaev, Sarosi, Rosenhaus, Trunin-SYK} (note the translation symmetry, which implies $G(\tau_1, \tau_2) = G(\tau_1 - \tau_2)$):
\beq \label{eq:SYK-propagator}
G(\tau) = G_0(\tau) + J^2 \int d\tau' d\tau'' \, G_0(\tau - \tau') \left[ G(\tau' - \tau'') \right]^{q-1} G(\tau''), \eeq
where $\tau = i t$ is the Euclidean time and $G_0(\tau) = \frac{1}{2} \, \mathrm{sgn}\!\left[ \sin\!\left( \pi \tau / \beta \right) \right]$ is the tree-level propagator at the inverse temperature $\beta$. The most remarkable property of this equation is its approximate conformal symmetry in the infrared limit. Namely, we expect on dimensional grounds that the solution to Eq.~\eqref{eq:SYK-propagator} decays with time as $G(\tau) \sim |\tau|^{-2/q}$. Hence, the l.h.s of Eq.~\eqref{eq:SYK-propagator} is negligible in the limit $J \beta \gg 1$ and $J \tau \gg 1$, so the equation is approximately invariant under orientation-preserving reparametrizations of the time circle, $\tau \to f(\tau)$, $f'(\tau) > 0$:
\beq \label{eq:SYK-reparametrizations}
G(\tau) \to G\left[ f(\tau) - f(0) \right] f(\tau)^\Delta f(0)^\Delta, \eeq
where the conformal dimension $\Delta = 1/q$. Such an invariance suggests the following ansatz on the resummed propagator:
\beq \label{eq:SYK-G-exact}
G(\tau) = B \left[ \frac{\pi}{J \beta \sin\left(\pi \tau/\beta\right)} \right]^{2\Delta} \mathrm{sgn}\!\left[ \sin\!\left( \frac{\pi \tau}{\beta} \right) \right]. \eeq
To determine the numerical coefficient $B = \left[ (1/2 - \Delta) \tan(\pi \Delta) / \pi \right]^\Delta$, we substitute the ansatz~\eqref{eq:SYK-G-exact} into the approximate Eq.~\eqref{eq:SYK-propagator}. Note that the solution~\eqref{eq:SYK-G-exact} indeed decays as $G(\tau) \sim |\tau|^{-2/q}$ in the limit $1 \ll J \tau \ll J \beta$, which confirms the self-consistency of the used approximation.

Moreover, in the limit $1 \ll q \ll N$, the solution to Eq.~\eqref{eq:SYK-propagator} can be represented as an expansion in $1/q$ near the tree-level propagator~\cite{Maldacena-SYK}:
\beq \label{eq:SYK-G-1/q}
G(\tau) \approx \frac{1}{2} \mathrm{sgn}\!\left[\sin\!\left( \frac{\pi \tau}{\beta} \right) \right] \times \left[ 1 + \frac{2}{q} \log \frac{\cos(\pi v/2)}{\cos\!\left( \pi v / 2 - \pi v |\tau|/\beta \right)} \right], \eeq
where the coefficient $v$ is determined from the equation
\beq \label{eq:SYK-v}
\beta \mathcal{J} = \pi v / \cos(\pi v / 2), \eeq
so $0 \le v < 1$, and the limit $v \to 1$ corresponds to the strong-coupling limit $\beta \mathcal{J} \gg 1$.

Finally, the R/A/K/W propagators in the SYK model are constructed using the fluctuation-dissipation theorem and an analytic continuation of the Euclidean propagator~\eqref{eq:SYK-G-exact} or~\eqref{eq:SYK-G-1/q}. Note that these propagators are defined on a $2n$-fold Keldysh contour, which corresponds to the effective temperature $\beta \to \beta_n = \beta (n+1)$, so we need to make the same change in Eqs.~\eqref{eq:SYK-G-exact}--\eqref{eq:SYK-v} before the analytic continuation. Thus, in the conformal limit, the R and W propagators are as follows: 
\beq \label{eq:SYK-G-exact-real} \begin{aligned}
i G^R(t) &= 2 B \cos(\pi \Delta) \left[ \frac{\pi}{J \beta_n \sinh\frac{\pi t}{\beta_n}} \right]^{2\Delta} \theta(t), \\
i G_{k,l}^W(t) &= B \left[ \frac{\pi}{J \beta_n \sinh\!\left( \frac{\pi t}{\beta_n} - \frac{i \pi (l-k)}{2 (n+1)} \right)} \right]^{2\Delta} \theta(t).
\end{aligned} \eeq
and in the limit $1 \ll q \ll N$, we have the following approximate identities:
\beq \label{eq:SYK-G-1/q-real} i G^R(t) \approx \theta(t), \qquad q J^2 \left[ i G_{k,l}^W(t) \right]^{q-2} \approx \frac{2 \pi^2 v_n^2}{\beta_n^2 \cosh^2\!\left[ \frac{\pi v_n t}{\beta_n} + \frac{i \pi v_n (n + 1 - l + k)}{2 (n+1)} \right]}. \eeq
The other two real-time propagators are restored in a similar way, but we do not need them to estimate the leading-order ROTOC and the refined quantum LE.

Now, let us calculate the regularized ROTOC, i.e., a fermionic analog of Eq.~\eqref{eq:ROTOC-reg}:
\beq \label{eq:SYK-ROTOC-def}
C_n^\mathrm{reg}(t) = \frac{1}{\tr \big( \hat{y}^{n+1} \big)} \tr\!\left[ \hat{y} \bigg( \frac{1}{N} \sum_{i,j} \hat{y}^{1/2} \left\{ \hat{\chi}_i(t), \hat{\chi}_j(0) \right\}^\dag \hat{y}^{1/2} \left\{ \hat{\chi}_i(t), \hat{\chi}_j(0) \right\} \bigg)^n \right], \eeq
where $\hat{y} = e^{- \beta \hat{H}}$ and we keep in mind that the canonical momenta of Majorana fermions are $\pi_i(t) = \chi_i(t)$, so the phase space has dimension $N$.

\begin{figure}[t]
    \center{\includegraphics[width=0.75\linewidth]{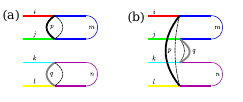}}
    \caption{Leading in $1/N$ (a) and subleading (b) loop corrections to $C_2^\mathrm{reg}(t)$ in the SYK model with $q = 3$. Different colors mark different chains of contracted indices. Namely, the left diagram is proportional to $\overline{j_{i,p,m} j_{j,p,m} j_{k,q,n} j_{l,q,n}} \sim \left( 1/N^4 \right) \delta_{i,j} \delta_{p,p} \delta_{m,m} \delta_{k,l} \delta_{q,q} \delta_{n,n} \sim N^4 / N^4 = \mO(1)$, and the right diagram  is proportional to $\overline{j_{i,p,m} j_{j,q,m} j_{k,q,n} j_{l,p,n}} \sim \left( 1/N^4 \right) \delta_{i,l} \delta_{p,p} \delta_{m,n} \delta_{j,k} \delta_{q,q} \delta_{m,n} \sim N^3 / N^4 = \mO(1/N)$.}
    \label{fig:SYK-factorization}
\end{figure}

Similarly to the bosonic vector mechanics (Sec.~\ref{sec:ON}), the tree-level ROTOC~\eqref{eq:SYK-ROTOC-def} equals a product of retarded propagators sitting on different folds, and the leading-in-$1/N$ loop corrections to the tree-level expression are given by ladder-type diagrams. However, unlike the vector mechanics, ladder diagrams  in the SYK model simply factorize (e.g., see Fig.~\ref{fig:SYK-factorization}) and can be resummed independently. 

For illustrative purposes, let us first consider the $n=2$ case. Similarly to Eqs.~\eqref{eq:ON-C2} and~\eqref{eq:ON-F}, we rewrite the ROTOC using an auxiliary function with several loose indices:
\beq \label{eq:SYK-C2}
C_2^\mathrm{reg}(t) = \sum_{i,i'} F_{i,i,i',i'}(t,0;t,0;t,0;t,0), \eeq
where $\mathbf{F} = F_{i_1,i_2,i_3,i_4}(t_1, t_1'; t_2, t_2'; t_3, t_3'; t_4, t_4')$ is as follows:
\beq \label{eq:SYK-F}
\mathbf{F} = \sum_{j, j'} \Big\langle \chi_{i_1; 1, c}(t_1) \, \chi_{j; 1, q}(t_1') \, \chi_{i_2; 2, c}(t_2) \, \chi_{j; 2, q}(t_2') \, \chi_{i_3; 3, c}(t_3) \, \chi_{j'; 3, q}(t_3') \, \chi_{i_4; 4, c}(t_4) \, \chi_{j'; 4, q}(t_4') \Big\rangle. \eeq

\begin{figure}[t]
    \center{\includegraphics[width=0.75\linewidth]{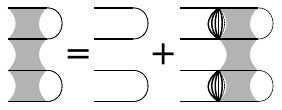}}
    \caption{The DS equation that sums the leading loop corrections to the function $\mathbf{F}$, which gives the ROTOC $C_2^\mathrm{reg}$ after index contractions, in the SYK model with $q = 7$. Horizontal lines correspond to the retarded propagators, and vertical lines correspond to the Wightman propagators. Thin lines denote index contractions where they do not follow the ordinary lines.}
    \label{fig:SYK-C2}
\end{figure}

Keeping in mind the factorization of ladder diagrams (Fig.~\ref{fig:SYK-factorization}) and resumming the melonic corrections to all propagators, we obtain an approximate (exact in the limit $N \to \infty$) DS equation on the function $\mathbf{F}$, see Fig.~\ref{fig:SYK-C2}:
\beq \label{eq:SYK-DS}
\mathbf{F} = \mathbf{F}_0 + \mathbf{K} \circ \mathbf{F}, \qquad \text{so} \qquad \mathbf{F} = (\mathbf{1} - \mathbf{K})^{-1} \circ \mathbf{F}_0. \eeq
Here, the action of an operator $\mathbf{K}$ is given by Eq.~\eqref{eq:ON-add-a-rung}, the zero-order contribution is equal to a product of retarded propagators:
\beq \label{eq:SYK-F_0}
\mathbf{F}_0 = \frac{1}{N^2} \delta_{i_1, i_2} \delta_{i_3, i_4} \tilde{G}^R(t_1, t_1') \tilde{G}^R(t_2, t_2') \tilde{G}^R(t_3, t_3') \tilde{G}^R(t_4, t_4'), \eeq
and the generating kernel $\mathbf{K}$ is represented as a tensor product of identical operators  acting on the corresponding invariant subspaces:
\beq \label{eq:SYK-K}
\mathbf{K} = \mK^{j_1, j_2}_{i_1, i_2}(t_1, t_1', t_2, t_2') \otimes \mK^{j_3, j_4}_{i_3, i_4}(t_3, t_3', t_4, t_4'), \eeq
where
\beq \label{eq:SYK-K-block}
\mK^{m,n}_{i,j}(t_1, t_1'; t_2, t_2') = J^2 (q-1) G^R(t_1, t_1') G^R(t_2, t_2') \left[ G^W_{k,k+1}(t_1', t_2') \right]^{q-1} \delta_{i,m} \delta_{j,n}. \eeq

In general, the operator product in the second Eq.~\eqref{eq:SYK-DS} can be estimated explicitly using the approximate conformal symmetry of the SYK model in the infrared limit. In that case, one determines the eigenfunctions and eigenvalues of Casimir operators and operators $\mK$ in each invariant subspace, constructs the eigenfunctions and eigenvalues of the total kernel $\mathbf{K} = \mK \otimes \mK$, projects $\mathbf{F}_0$ onto this basis, and explicitly calculates the sums over all eigenfunctions. 

Unfortunately, these calculations are very tedious, although straightforward. So, for brevity, let us instead qualitatively estimate the LOTOC and the replica LEs in the limit $1 \ll q \ll N$. To that end, we substitute the exponentially growing ansatz~\eqref{eq:ON-ansatz} to the first Eq.~\eqref{eq:SYK-DS}, neglect the bounded function $\mathbf{F}_0$, take into account approximate identities~\eqref{eq:SYK-G-1/q-real} and factorization Fig.~\ref{fig:SYK-factorization}, and calculate all time integrals. As a result, we obtain the following equation on $\varkappa_2$:
\beq \label{eq:SYK-replica-2}
\mathbf{F} \approx \mK^{\otimes 2} \circ \mathbf{F} \Longrightarrow \mathbf{F} \approx \left( \frac{2 \pi v_2}{\beta_2 \varkappa_2} \right)^2 \mathbf{F} \Longrightarrow 1 \approx \left( \frac{2 \pi v_2}{\beta_2 \varkappa_2} \right)^2. \eeq
We emphasize that unlike Eqs.~\eqref{eq:ON-replica-LE-eq} and~\eqref{eq:ON-replica-LE-eq-truncated}, this equation has no combinatorial factor due to factorization of ladder diagrams.

Generalizing Eq.~\eqref{eq:SYK-replica-2} to arbitrary replica numbers:
\beq \label{eq:SYK-replica-n}
1 \approx \left( \frac{2 \pi v_n}{\beta_n \varkappa_n} \right)^n. \eeq
and keeping in mind the corresponding ansatz~\eqref{eq:ON-ansatz-n}, we estimate the replica LE $\kappa_n = n \varkappa_n$:
\beq \label{eq:SYK-replica-LE}
\kappa_n \approx \frac{2 \pi n v_n}{\beta_n}. \eeq
Finally, we estimate the LOTOC and the refined quantum LE employing the replica trick~\eqref{eq:replica-trick}:
\beq \label{eq:SYK-LOTOC}
L(t) \approx 2 \bar{\kappa}_q t + \const \qquad \text{and} \qquad \bar{\kappa}_q \approx \frac{2 \pi v}{\beta}, \eeq
where the coefficient $v$ is determined from Eq.~\eqref{eq:SYK-v}. The constant contribution to the LOTOC and the subleading corrections to the refined quantum LE, as well as analogs of Eq.~\eqref{eq:SYK-LOTOC} for arbitrary~$q$, can be calculated numerically or analytically using the approximate conformal symmetry of the SYK model, e.g., see~\cite{Maldacena-SYK, Kitaev} and~\cite{Romero-Bermudez} for $n=1$ case. 

We emphasize that the refined quantum LE and the conventional quantum LE coincide in the SYK model because the correlations between different replicas are suppressed by powers of $1/N$. In particular, the refined quantum LE satisfies the bound on chaos, $\bar{\kappa}_q \le 2\pi/\beta$, which was proven for conventional quantum LEs of the large-$N$ models~\cite{MSS}. Furthermore, the refined quantum LE saturates this bound in the strong-coupling limit $\beta \mathcal{J} \gg 1$. This supports the putative duality of the SYK model and the Jackiw-Teitelboim gravity in a near-AdS$_2$ space, which models an extremal black hole and therefore must satisfy the bound~\cite{MSS} for \textit{both} conventional and refined quantum LEs to avoid the information cloning paradox~\cite{Hayden, Sekino}.

\subsection{Matrix field theory at weak coupling}
\label{sec:matrix}

To illustrate the calculation of the ROTOC and the refined quantum LE in a quantum field theory, we consider a Hermitian matrix field $\Phi_{ab}$ with a quartic interaction following~\cite{Stanford:phi-4, Grozdanov, Romero-Bermudez}:
\beq \label{eq:H-matrix}
H = \int d^3 x \, \tr\!\left[ \sum_{a,b = 1}^N \left( \frac{1}{2} \dot{\Phi}_{ab} \dot{\Phi}_{ba} + \frac{1}{2} \mathbf{\nabla} \Phi_{ab} \mathbf{\nabla} \Phi_{ba} + \frac{m^2}{2} \Phi_{ab} \Phi_{ba} \right) + \frac{\lambda}{4 N} \sum_{a,b,c,d = 1}^N \Phi_{ab} \Phi_{bc} \Phi_{cd} \Phi_{da} \right]. \eeq
For simplicity, we assume that the coupling constant is small, $\lambda \ll 1$, and estimate the refined quantum LE only to the leading order in $\lambda$ and $1/N$.

Similarly to the vector mechanics, Sec.~\ref{sec:ON}, we introduce the retarded, advanced, Keldysh, and Wightman propagators of the matrix fields on a $2n$-fold Keldysh contour:
\beq \label{eq:matrix-propagators-def} \begin{aligned}
i G_{a,b; a',b'}^R(t, t') &= i G^R(t, t') \delta_{a,a'} \delta_{b,b'} = \left\langle \hat{\Phi}_{a,b | k,c}(t) \hat{\Phi}_{a',b' | k,q}(t') \right\rangle, \\
i G_{a,b; a',b'}^A(t, t') &= i G^A(t, t') \delta_{a,a'} \delta_{b,b'} = \left\langle \hat{\Phi}_{a,b | k,q}(t) \hat{\Phi}_{a',b' | k,c}(t') \right\rangle, \\
i G_{a,b; a',b'}^K(t, t') &= i G^K(t, t') \delta_{a,a'} \delta_{b,b'} = \left\langle \hat{\Phi}_{a,b | k,c}(t) \hat{\Phi}_{a',b' | k,c}(t') \right\rangle, \\
i G_{a,b; a',b' | k,l}^W(t, t') &= i G_{k,l}^W(t, t') \delta_{a,a'} \delta_{b,b'} = \left\langle \hat{\Phi}_{a,b | k,c}(t) \hat{\Phi}_{a',b' | l,c}(t') \right\rangle,
\end{aligned} \eeq
where $a,b,a',b'$ denote the group indices, $k,l$ numerate the folds, and $c,q$ mark the classical and quantum components of the field. At the tree level, the propagators~\eqref{eq:matrix-propagators-def} have the following form:
\beq \label{eq:matrix-propagators-tree} \begin{aligned}
i G^R(k) &= \frac{i}{2 E_\mathbf{k}} \left( \frac{1}{k_0 - E_\mathbf{k} + i0} - \frac{1}{k_0 + E_\mathbf{k} + i0} \right), \\
i G^A(k) &= \frac{i}{2 E_\mathbf{k}} \left( \frac{1}{k_0 - E_\mathbf{k} - i0} - \frac{1}{k_0 + E_\mathbf{k} - i0} \right), \\
i G^K(k) &= \frac{1}{2} \coth \!\left[ \frac{\beta (n+1) k_0}{2} \right] \frac{\pi}{E_\mathbf{k}} \Big[ \delta\!\left(k_0 - E_\mathbf{k} \right) - \delta\!\left(k_0 + E_\mathbf{k} \right) \Big], \\
i G^W_{k,l}(k) &= \frac{e^{\beta (l - k) k_0}}{e^{\beta (n + 1) k_0} - 1} \, \frac{\pi}{E_\mathbf{k}} \Big[ \delta\!\left(k_0 - E_\mathbf{k} \right) - \delta\!\left(k_0 + E_\mathbf{k} \right) \Big].
\end{aligned} \eeq
Here, we use the symmetry of the model~\eqref{eq:H-matrix} under time and space translations, which allows us to do a Fourier transform of the propagators. Obviously, $k = (k_0, \mathbf{k})$ and $E_\mathbf{k} = \sqrt{m^2 + \mathbf{k}^2}$. The initial state is a thermal state with inverse temperature $\beta_n = (n+1) \beta$, so the R/A/K propagators are related by the fluctuation-dissipation theorem. An arbitrary Wightman propagator is constructed from a Wightman propagator $i G^W_{1,1}(k)$ by translations in imaginary time, see~\cite{Romero-Bermudez}.

The vertices in the model~\eqref{eq:H-matrix} coincide with the vertices depicted at Fig.~\ref{fig:diagram-rules}. The only difference is that now, all lines carry two group indices, so the standard graphs are replaced with ribbon graphs.

At the tree level, the ROTOCs in the model~\eqref{eq:H-matrix} break down in the product of retarded propagators similarly to Eqs.~\eqref{eq:ROTOC-phi-4-tree} and~\eqref{eq:ON-F_0}. Then, similarly to the vector mechanics (Sec.~\ref{sec:ON}), this tree-level correlator receives two types of loop corrections.

First, loop corrections to the retarded propagators renormalize the corresponding self-energies. In quantum mechanics, such corrections resulted in a simple shift of the mass~\eqref{eq:ON-m}. However, in a quantum field theory, the loop-corrected self-energy also acquires an imaginary contribution (the simplest one comes from the two-loop ``sunset'' diagram), which implies an exponential decay of correlation functions\footnote{A quantum mechanical model with a finite number of degrees of freedom cannot thermalize without an external bath because its spectrum is discrete, and correlation functions bound to oscillate with some (possibly large) period. At the same time, a quantum field comprises infinitely many degrees of freedom and serves as a bath to itself.}, i.e., replaces $E_\mathbf{k} \pm i 0$ by $E_\mathbf{k} \pm i \Gamma_\mathbf{k}$ in the tree-level propagators~\eqref{eq:matrix-propagators-tree}. At the two-loop order, this imaginary contribution has the following form:
\beq \label{eq:matrix-gamma}
\Gamma_\mathbf{p} \approx \frac{1}{6} \int \frac{d^3 \mathbf{q}}{(2 \pi)^3} \frac{e^{\beta(l-k) q_0}}{e^{\beta (n+1) q_0} - 1} \frac{e^{\beta (n+1) p_0} - 1}{e^{\beta(l-k) p_0}} \frac{R_{k,l}(E_\mathbf{q} + E_\mathbf{p}, \mathbf{q} + \mathbf{p}) + R_{k,l}(E_\mathbf{q} - E_\mathbf{p}, \mathbf{q} - \mathbf{p})}{E_\mathbf{p} E_\mathbf{q}}, \eeq
where $R_{k,l}(p)$ is the rung function that connects two different folds similarly to vertical parts of building blocks $\mK^1_{k,l}$ and $\mK^2_{k,l}$ from Sec.~\ref{sec:ON}:
\beq \label{eq:matrix-R}
R_{k,l}(p) = 3 \lambda^2 \int \frac{d^4 q}{(2 \pi)^4} G_{k,l}^W(p/2 + q) G_{k,l}^W(p/2 - q). \eeq
Of course, the integral in Eq.~\eqref{eq:matrix-gamma} does not depend on the fold indices $k$ and $l$; we just rewrite it in this form for convenience.

\begin{figure}[t]
    \center{\includegraphics[width=0.75\linewidth]{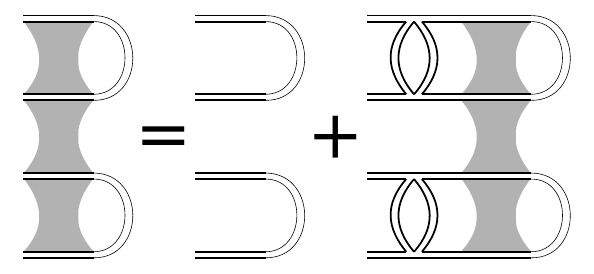}}
    \caption{The DS equation that sums the leading in $\lambda$ and $1/N$ corrections to the tree-level ROTOC $C_2^\mathrm{reg}$ in the model~\eqref{eq:H-matrix}. Horizontal ribbons correspond to the retarded propagators, and vertical ribbons correspond to the Wightman propagators. Thin lines indicate index contractions where they do not follow ordinary lines.}
    \label{fig:matrix-DS}
\end{figure}

Second, there are loop corrections to the tree-level ROTOC that connect different folds via rung functions~\eqref{eq:matrix-R}, see Fig.~\ref{fig:matrix-DS}. In fact, the leading-order-in-$1/N$ corrections connect only the neighboring folds because only planar diagrams do survive in the limit $N \to \infty$. Furthermore, it is straightforward to see that in the planar limit, the ROTOCs on a $2n$-fold contour factorize into a product of independent sums of ladder diagrams, i.e., a product of OTOCs on a $2n$-fold contour:
\beq \label{eq:matrix-ROTOC}
C_n^\mathrm{reg}(t) = \tilde{C}(t)^n = \left[ \frac{1}{N^2} \sum_{i,j=1}^2 \sum_{a,b,a',b'} \frac{1}{\tr \big( \hat{y}^{n+1} \big)} \tr\!\left( \hat{y}^{n + 1/2} \left[ \hat{z}_{a,b}^i(t), \hat{z}_{a',b'}^j(0) \right]^\dag \hat{y}^{1/2} \left[ \hat{z}_{a,b}^i(t), \hat{z}_{a',b'}^j(0) \right] \right) \right]^n, \eeq
where $z_{a,b}^1(t) = \Phi_{a,b}(t, \mathbf{x})$, $z_{a,b}^2(t) = \pi_{a,b}(t, \mathbf{x}) = m \dot{\Phi}_{a,b}(t, \mathbf{x})$, and $\hat{y} = e^{-\beta \hat{H}}$. In other words, the correlations between different folds are suppressed by powers of $1/N$ similarly to the SYK model.

Due to factorization, the DS equations on each OTOC can be solved independently and are straightforwardly reduced to the following equation on an auxiliary function $f(\omega, \mathbf{p})$:
\beq \label{eq:matrix-DS-eq}
- i \omega f(\omega, \mathbf{p}) \approx -2 \Gamma_\mathbf{p} f(\omega, \mathbf{p}) + \int\frac{d^3 \mathbf{q}}{(2 \pi)^3} \frac{R_{k,k+1}(E_\mathbf{q} + E_\mathbf{p}, \mathbf{q} + \mathbf{p}) + R_{k,k+1}(E_\mathbf{q} - E_\mathbf{p}, \mathbf{q} - \mathbf{p})}{4 E_\mathbf{p} E_\mathbf{q}} f(\omega, \mathbf{q}), \eeq
such that
\beq \label{eq:matrix-OTOC-Fourier}
\tilde{C}(t) \approx \int \frac{d \omega}{2 \pi} e^{-i \omega t} \left( 1 + m^2 \omega^2 \right) \int \frac{d^4 p}{(2 \pi)^4} \, f(\omega, \mathbf{p}) \delta\!\left(p_0^2 - E_\mathbf{p}^2 \right). \eeq
After inverse partial Fourier transform over $\omega$, we obtain an integro-differential equation of the following form:
\beq \frac{\pd}{\pd t} f(t, \mathbf{p}) \approx \mathcal{M}_n(\mathbf{p}, \mathbf{q}) \circ f(t, \mathbf{q}), \eeq
where the integral operator $\mathcal{M}_n(\mathbf{p}, \mathbf{q})$ is derived from the r.h.s of Eq.~\eqref{eq:matrix-DS-eq}. The eigenvalues $\varkappa_n$ of this operator can be found numerically following~\cite{Stanford:phi-4, Grozdanov, Romero-Bermudez}. In particular, the largest positive eigenvalue is estimated as follows in the high-temperature:
\beq \label{eq:matrix-kappa-high}
\varkappa_n \approx 0.0015 \frac{\lambda^2}{m \beta_n^2} \qquad \text{as} \qquad \beta m \ll 1, \eeq
and low-temperature limit:
\beq \label{eq:matrix-kappa-low}
\varkappa_n \approx 4 \, \frac{\lambda^2}{m \beta_n^2} \, e^{-\beta (2 n + 1) m} \qquad \text{as} \qquad \beta m \gg 1. \eeq
Keeping in mind the factorization of the ROTOCs, we estimate the replica LE $\kappa_n = n \varkappa_n / 2$ and the refined quantum LE in the model~\eqref{eq:H-matrix}:
\beq \label{eq:matrix-true-qLE}
\bar{\kappa}_q \approx \begin{cases} 0.001 \lambda^2 / m \beta^2 \quad &\text{as} \quad \beta m \ll 1, \\ 2 \left(\lambda^2 / m \beta^2\right) e^{-\beta m} \quad &\text{as} \quad \beta m \gg 1. \end{cases} \eeq
Note that this behavior is qualitatively similar to the chaotic vector mechanics~\eqref{eq:H-ON}, where the refined quantum LE always remain small, grows as a power of temperature at high temperatures, and is exponentially suppressed at small temperatures, cf. Eqs.~\eqref{eq:ON-true-qLE}--\eqref{eq:ON-true-qLE-simplified-low}. However, we emphasize that unlike the vector mechanics, where correlations between different replicas lead to an extra numerical factor in front of the refined quantum LE, correlations between different replicas are suppressed by powers of $1/N$ in the model~\eqref{eq:H-matrix}, so its refined and conventional quantum LEs coincide.

\subsection{Regularization dependence of replica OTOCs}
\label{sec:regularization}

Until the present moment, we ignored a subtlety in the definitions of the regularized ROTOC~\eqref{eq:ROTOC-reg} and LOTOC~\eqref{eq:LOTOC-reg}: in general, we can regularize~\eqref{eq:LOTOC-def} by shifting times in $\hat{z}_i(t) \to \hat{z}_i(t + i \beta \eta)$ and $\hat{z}_i(0) \to \hat{z}_i(i \beta \xi)$ with arbitrary $0 < \xi < \eta$. For simplicity, we chose the symmetric regularization with $\xi = \eta = 1/4$, in which all folds of the contour (Fig.~\ref{fig:contour}) are equidistant, but there is no evident reason for such a choice. Worse than that, the ROTOCs and the LOTOC might depend on~$\xi$ and~$\eta$. In fact, the contour dependence of the ROTOCs becomes evident already at the $n = 1$ level, i.e., for ordinary OTOCs~\cite{Liao, Romero-Bermudez, Sahu}. In this subsection, we argue that the refined quantum LEs~\eqref{eq:ON-true-qLE} and~\eqref{eq:SYK-LOTOC} do not depend on the regularization, whereas the refined quantum LE~\eqref{eq:matrix-true-qLE} does.

In the vector mechanics, Sec.~\ref{sec:ON}, the leading nontrivial loop corrections to the ROTOC $C_n^\mathrm{reg}(t)$ are generated by the kernel $\mathbf{K}$, which consists of building blocks $\mK^1$ and $\mK^2$. In general, the eigenvalues of these building blocks depend on $\xi$ and $\eta$. However, as we pointed in Sec.~\ref{sec:ON}, the \textit{largest real part} of all eigenvalues does not depend on where the real folds are inserted in the Keldysh contour (as long as its imaginary part has a fixed total length). The same conclusion applies to the largest real eigenvalue of the kernel $\mathbf{K}$, since this eigenvalue equals a simple sum of the eigenvalues of the corresponding building blocks. This readily implies that the replica LEs and the refined quantum LE in the model~\eqref{eq:H-ON} do not depend on $\xi$ and $\eta$. 

In the double-scaled SYK model ($1 \ll q \ll N$), Sec.~\ref{sec:SYK}, the only place where $\xi$ and $\eta$ enter the correlation functions is the imaginary argument of hyperbolic cosine in Eq.~\eqref{eq:SYK-G-1/q-real}. Hence, following~\cite{Romero-Bermudez}, we can reabsorb the contour dependence of $\mK$ and $\mathbf{K}$ with a redefinition of $\mathbf{F}$ in Eq.~\eqref{eq:SYK-replica-2}. A similar reasoning also applies to a finite-$q$ version of the SYK model. However, in the double-scaled SYK model, there is an alternative way to see that the refined quantum LE does not depend on the regularization. Following~\cite{Maldacena-SYK}, we can reduce the eigenvalue problem of~$\mK$ to a single-particle Schr{\"o}dinger equation in a complex P{\"o}schl-Teller potential, find a single bound state of this equation, and explicitly show that the corresponding energy, which is proportional to a real eigenvalue of $\mK$, does not depend on $\xi$ and $\eta$. In other words, eigenvalues of $\mK$ and $\mathbf{K}$ are contour independent, and so are the replica LEs and the refined quantum LE.

In fact, the contour independence of the quantum LEs in the vector and SYK models is related to their gapless nature (approximate at finite $N$, but exact in the large-$N$ limit). First, we expect that the building blocks $\mK$, $\mK^{1}$, and $\mK^2$, which describe the scattering of excitations in the corresponding models, are peaked around the value of the gap in the frequency space. Second, the contour dependence of these building blocks reduces to a single prefactor of $e^{\beta (\omega - \omega') (\xi + \eta - 1) / 2}$, which equals unity at the origin and exponentially decays away from it~\cite{Romero-Bermudez}. Third, the size of the gap is exponentially small in the SYK model~\cite{Maldacena-SYK} and proportional\footnote{The $O(N)$-symmetric version of~\eqref{eq:H-ON} has a gap due to finite mass of excitations, and each level is $N$-times degenerate. However, the non-symmetric term lifts the degeneracy and makes the spectrum almost equidistant.} to $1/N^2$ in the vector mechanics~\eqref{eq:H-ON}. In other words, the building blocks and the kernels that generate loop corrections in both these models are peaked around the origin in the large-$N$ limit. Therefore, in both these models, the contour-dependent factor approximately equals unity for arbitrary~$\xi$ and~$\eta$, so the replica LEs and the refined quantum LEs are contour independent. 

In the matrix field theory, Sec.~\ref{sec:matrix}, the situation is different. This model is massive and thus clearly has a gap. Applying the reasoning of the previous paragraph to a gapped model, we find that the eigenvalues of the generating kernel and, hence, the replica LEs exponentially depend on~$\xi$ and~$\eta$. A numerical calculation confirms this qualitative estimate, e.g., see~\cite{Romero-Bermudez} for the estimates in the $n=1$ case (the factorization of ladder diagrams implies a straightforward extension of this result to arbitrary $n$). However, when the temperature is significantly higher than the size of the gap, this dependence is negligible, so the approximate identities~\eqref{eq:matrix-kappa-high} and $\bar{\kappa}_q \approx 0.001 \lambda^2 / m \beta^2$ as $\beta m \ll 1$ still hold for arbitrary~$\xi$ and~$\eta$. For the same reasons, we expect that the high-temperature behavior of the refined quantum LE does not depend on regularization in an arbitrary chaotic model.

\section{Discussion and Conclusion}
\label{sec:discussion}

In this paper, we analytically calculated the refined quantum LEs using the replica trick~\eqref{eq:replica-trick} in a variety of quantum chaotic models. In particular, we confirmed the correspondence between the refined quantum LE and the average classical LE in simple quantum mechanical examples of harmonic oscillator, LMG model, and quantum cat map. We also suggested a generalization of the Schwinger-Keldysh diagram technique that is suitable for calculations of arbitrary ROTOCs and replica LEs and is especially useful for large-$N$ models. To illustrate this technique, we considered three paradigmatic examples of large-$N$ models: \textit{vector} mechanics, \textit{matrix} field theory, and the SYK model, which can be reformulated as a \textit{tensor} model, see~\cite{Klebanov-1, Klebanov-2, Klebanov-3}.

Furthermore, we directly observed how correlations between different replicas affect the refined quantum LE. In the matrix and the SYK model, the correlations between replicas are suppressed by the powers of $1/N$, so the refined and conventional quantum LEs coincide. However, in the vector mechanics, correlations are not suppressed, assign a nontrivial factor to the replica LEs, and reduce the refined quantum LE comparing to the conventional one. This observation motivates a more careful study of quantum LEs in chaotic quantum systems --- especially in the systems that saturate the bound $\kappa_q \le 2 \pi T / \hbar$ and are supposed to be dual to a black hole~\cite{MSS}.

However, we emphasize that the LOTOC preserves the most important property of the quantum LE --- the bound on chaos~\cite{MSS}, which is extremely important for information and cloning paradoxes~\cite{Hayden, Sekino}. On the one hand, the semiclassical analysis of~\cite{companion}, supported by the large-$N$ calculations of Sec.~\ref{sec:diagrams}, implies that the refined quantum LE is always smaller (more precisely, not greater) than the conventional one\footnote{The OTOC singles out only the points with the largest LEs, whereas the LOTOC calculates a fair average of all LEs, so the refined quantum LE cannot be greater than the conventional one by definition.}, $\bar{\kappa}_q \le \kappa_q$. On the other hand, replica LEs are proven~\cite{Pappalardi, Tsuji} to satisfy the bound $\kappa_n \le 2 \pi n T / \hbar$ for any positive integer $n$, which reproduces the bound~\cite{MSS} on the refined quantum LE after analytic continuation to real $n$ and use of the replica trick~\eqref{eq:replica-trick}. Thus, the refined quantum LE and scrambling time satisfy the same bounds as their conventional versions extracted from the OTOC, i.e., $\bar{\kappa}_q \le 2 \pi T / \hbar$ and $t_E \ge \left( 1 / \bar{\kappa}_q \right) \log(N)$ if $\bar{\kappa}_q > 0$.

The approaches of the present work can be extended in many directions.

First, the definition of the refined quantum LE~\eqref{eq:true-qLE-def} can be extended to the full quantum Lyapunov spectrum. To that end, we simply remove the sum and introduce a quantum analog of classical matrix $\Lambda_{ij}$, cf.~\eqref{eq:cl-LE-spectrum}: 
\beq \label{eq:q-LE-spectrum}\left\langle \log\!\bigg( \frac{1}{N \hbar^2} \left[ \hat{z}_i(t), \hat{z}_j(0) \right]^\dag \left[ \hat{z}_i(t), \hat{z}_j(0) \right] \bigg) \right\rangle = 2 \tilde{\Lambda}_{ij} t + o(t) \quad \text{as} \quad 1 \ll t \ll t_E. \eeq
Diagonalizing matrix $\tilde{\Lambda}_{ij}$, we determine all $N$ quantum LEs, which are expected to reproduce the corresponding average classical LEs in the limit $\hbar \to 0$. Furthermore, using such a quantum Lyapunov spectrum, we can calculate a quantum analog of the Kolmogorov-Sinai entropy similarly to~\eqref{eq:KS-entropy}. It is natural to expect that the information scrambling is related to the Kolmogorov-Sinai entropy, which measures a gross exponential divergence of classical trajectories in all directions, rather than the maximal LE. Therefore, it is promising to calculate this entropy in black holes and other prominent quantum chaotic systems.

Moreover, it would be interesting to compare our definition of quantum Kolmogorov-Sinai entropy to other known definitions, including dynamical entropy~\cite{Connes, Alicki}, entanglement-based~\cite{Bianchi}, and OTOC-based quantities~\cite{Gharibyan:2018, Lerose, Correale}. The advantage of our approach is a clear correspondence between Eq.~\eqref{eq:q-LE-spectrum} and the classical Kolmogorov-Sinai entropy, which establishes a direct bridge between quantum chaos and classical K-systems.

Second, the LOTOC~\eqref{eq:LOTOC-def} can be used to measure the operator growth in spatially extended chaotic quantum systems, such as spin chains, random circuits, or quantum field theories~\cite{Swingle:2016, Roberts:2016, Nahum, Mi, Xu:2018, Xu-tutorial}. Do LOTOCs capture any details of the operator growth that OTOCs miss? Do the butterfly velocities defined using OTOCs and LOTOCs coincide? Are there spatially extended analogs of integrable systems with unstable fixed points, where OTOCs exponentially grow, and LOTOCs do not? In particular, how do LOTOCs respond to quantum scars? We believe that all these questions are closely related to thermalization of quantum systems and, hence, deserve careful study.

Third, it is promising to extend the universal approaches~\cite{Gu, Stanford:2021, Choi:2023, Blake, Gao} from the OTOCs to the ROTOCs and the LOTOCs. In particular, we believe that such an effective picture, where scrambling is associated with exchanges of ``scramblons'' or ``hydrodynamic modes'', will shed some light on the nature of correlations between different replicas and help us to estimate the ROTOCs analytically in a much wider class of chaotic systems.

Fourth, it is interesting to study the semiclassical limit of the LOTOC using the methods of~\cite{Cotler:2017, Rammensee, Jalabert:2018}. This will give a rigorous proof of the correspondence between the refined quantum LE and the average classical LE instead of our qualitative reasoning and case studies.

Fifth, the breakdown of the factorization ansatz for the ROTOC ($C_n(t) = C_1(t)^n$), which we observed in the vector mechanics (Sec.~\ref{sec:ON}), resembles the replica symmetry breaking in spin glasses~\cite{Mezard, Castellani}. Moreover, we believe that the connection between these two phenomena might be deeper than a superficial similarity of two techniques. On the one hand, the correlations between replicas reflect the overlaps between different pure states in the original model, which, in turn, reflect the structure of its phase space. On the other hand, the LOTOC describes the semiclassical dynamics on the same phase space (up to the Ehrenfest time). Therefore, it seems promising to study the relation between replica symmetry breaking and divergence of refined and conventional quantum LEs. As a convenient example, one can consider the non-Hermitian SYK model, where replica symmetry breaking solutions dominate and correlation functions can be calculated explicitly~\cite{nH-SYK-1, nH-SYK-2}.

Finally, it would be very useful to apply the AdS/CFT correspondence~\cite{Maldacena-holography, Aharony, Gubser, Witten:1998} to calculations of the ROTOCs and LOTOCs. To that end, we need to generalize the standard holographic approach, which works with Euclidean or Schwinger-Keldysh contours~\cite{Skenderis-1, Skenderis-2, Liu:2018, deBoer}, to a multi-fold Keldysh contour. On the one hand, such an extended holographic Schwinger-Keldysh approach should help us to determine the refined quantum LE of a black hole, pure anti-de~Sitter space, a static patch of the de~Sitter space, or the corresponding dual conformal field theories. On the other hand, it might deepen our understanding of black holes. For example, it is extremely interesting to check whether replica wormholes contribute to the ROTOCs and the LOTOC. Indeed, the LOTOC is nothing but a dynamical generalization of the entropy, cf. Eq.~\eqref{eq:LOTOC-reg}, and replica wormholes do leave footprints in the entropy of a black hole~\cite{Penington, Almheiri}. Therefore, we expect a similar consequence for the LOTOC and the refined quantum LE (see also the discussion of replica contributions to multipoint correlation functions in the semiclassical gravity~\cite{Stanford:2020}).

\section*{Acknowledgments}
We thank Nikita Kolganov, Artem Alexandrov, Tigran Sedrakyan, Anatoly Dymarsky, Silvia Pappalardi, Ekaterina Izotova, Elizaveta Trunina, Damir Sadekov, Pavel Orlov, and Emil Akhmedov for valuable discussions. This work was supported by the grant from the Foundation for the Advancement of Theoretical Physics and Mathematics ``BASIS''.

%

\end{document}